\documentclass[conference]{IEEEtran}
\usepackage{cite}
\usepackage{amsmath,amssymb,amsfonts}
\usepackage{mathptmx} 
\usepackage{comment}
\usepackage{bm}
\usepackage{algorithmic}
\usepackage{graphicx}
\usepackage{textcomp}
\usepackage{fancyhdr}
\usepackage[hyphens]{url}
\usepackage[normalem]{ulem}
\usepackage[bookmarks=true,breaklinks=true,colorlinks,linkcolor=black,citecolor=blue,urlcolor=black]{hyperref}
\hypersetup{
    unicode=false,          % non-Latin characters in Acrobat’s bookmarks
    pdftoolbar=true,        % show Acrobat’s toolbar?
    pdfmenubar=true,        % show Acrobat’s menu?
    pdffitwindow=false,     % window fit to page when opened
    pdftitle={Booster},    % title
    pdfauthor={Mithuna Thottethodi},     % author
    pdfsubject={Booster},   % subject of the document
    pdfcreator={Mithuna Thottethodi},   % creator of the document
    pdfproducer={},         % producer of the document
    pdfkeywords={Gradient Boosting,} {Booster}, % list of keywords
    pdfnewwindow=true,      % links in new window
    colorlinks=false,       % false: boxed links; true: colored links
    linkcolor=red,          % color of internal links
    citecolor=green,        % color of links to bibliography
    filecolor=magenta,      % color of file links=true,        
    urlcolor=cyan           % color of external links
}

\def\BibTeX{{\rm B\kern-.05em{\sc i\kern-.025em b}\kern-.08em
    T\kern-.1667em\lower.7ex\hbox{E}\kern-.125emX}}

\newcommand{\step}[1]{{Step} \raisebox{.5pt}{\textcircled{\raisebox{-.9pt} {#1}}}\xspace}
\newcommand{\rawstep}[1]{\raisebox{.5pt}{\textcircled{\raisebox{-.9pt} {#1}}}\xspace}

\newcommand{\ig}{{\em Ideal GPU}\xspace}
\newcommand{\im}{{\em Ideal 32-core}\xspace}
\newcommand{\flight}{{\tt Flight}\xspace}
\newcommand{\allstate}{{\tt Allstate}\xspace}
\newcommand{\higgs}{{\tt Higgs}\xspace}
\newcommand{\iot}{{\tt IoT}\xspace}
\newcommand{\mq}{{\tt Mq2008}\xspace}

\title{Booster: An Accelerator for Gradient Boosting Decision Trees}
\author{Mingxuan He, T. N. Vijaykumar, and Mithuna Thottethodi\\
School of Electrical and Computer Engineering, Purdue University  \\
Email: {\tt he238@purdue.edu, vijay@ecn.purdue.edu, mithuna@purdue.edu}
}

\usepackage{graphicx}
\usepackage{ifthen}
\usepackage{calc}
\usepackage{pifont}
\usepackage{soul}
\usepackage{fancyhdr}
\usepackage{xspace}
\usepackage[table]{xcolor}
\definecolor{lightgray}{gray}{0.9}
\newenvironment{stripetabular}{\rowcolors{2}{white}{lightgray}\tabular}{\endtabular}

\newcounter{hours}
\newcounter{minutes}

\definecolor{brightgreen}{rgb}{0.4, 1.0, 0.0}
\sethlcolor{brightgreen}

\newcommand{\name}{{Booster}\xspace}

\newboolean{timeofmake}
\setboolean{timeofmake}{false}

\newcommand{\ignore}[1]{}

\newcommand{\dontinclude}[1]{ }

\newcommand{\putsec}[2]{\vspace{-0.0in}\section{#2}\label{sec:#1}\vspace{-0.0in}}
\newcommand{\putsubsec}[2]{\vspace{0.0in}\subsection{#2}\label{sec:#1}\vspace{0.00in}}

\newcommand{\putsubsechl}[2]{\vspace{0.0in}\subsection{#2}\label{sec:#1}\vspace{0.00in}}

\newcommand{\tabput}[3]{
\begin{table}[t]
\footnotesize 
\caption{\footnotesize\sf #3 \label{tab:#1}}
\vspace{-0.15in}
\begin{center}
{
#2
}
\end{center}
\vspace{-0.25in}
\end{table}
}

\newcommand{\figput}[4][1.0\linewidth]{
\begin{figure}[t]
\begin{minipage}{\linewidth}
\footnotesize 
\begin{center}
\includegraphics[clip, width=#1]{figures/#2}
\end{center}
\vspace{-0.22in}
\caption{#4 \label{fig:#2}}
\vspace{-0.15in}
\end{minipage}
\end{figure}
}

\newcommand{\figref}[1]{Figure~\ref{fig:#1}}
\newcommand{\tabref}[1]{Table~\ref{tab:#1}}
\newcommand{\secref}[1]{Section~\ref{sec:#1}}

\begin{document}
\maketitle
\thispagestyle{plain}
\pagestyle{plain}

\begin{abstract}
Recent breakthroughs in machine learning (ML) have 
sparked hardware innovation for efficient execution of these emerging ML workloads.
Separately, due to  recent refinements and high-performance implementations, well-established  {\em gradient boosting decision tree} models (e.g., XGBoost) have demonstrated their dominance in many real-world applications.
Beyond its rich theoretical foundations, gradient boosting is prevalent in commercially-important contexts, such as table-based datasets (e.g., those held in relational  databases and spreadsheets). 
While not as computationally intensive as the training of  deep, convolutional neural networks,  gradient boosting training is time-consuming  (several hours for large  datasets).
Despite their importance, gradient boosting models have not been targeted much for hardware acceleration. 

We propose \name, a novel accelerator for gradient boosting trees based on the unique characteristics of gradient boosting models. We observe that the dominant steps of gradient boosting training (accounting for 90-98\% of training time) 
involve simple, fine-grained, independent operations on small-footprint data structures (e.g., accumulate and compare values in the structures). 
Unfortunately, existing multicores and GPUs are unable to harness this parallelism 
because they do not support massively-parallel data structure accesses that are irregular and data-dependent. 
By employing a {\em scalable sea-of-small-SRAMs} approach and an SRAM bandwidth-preserving mapping of data record fields to the SRAMs, 
\name  achieves  significantly more parallelism (e.g., 3200-way parallelism) than multicores and GPU. In addition,  \name employs 
a redundant data representation that significantly lowers the memory bandwidth demand. 
Our simulations reveal that 
\name achieves 11.4x speedup and 6.4x speedup over an ideal 32-core multicore and an ideal GPU, respectively. Based on ASIC synthesis of FPGA-validated RTL using 45 nm technology, we estimate a \name chip to occupy 60 mm$^2$ of area and  dissipate 23 W when operating at  1-GHz clock speed.  
\end{abstract}

\putsec{intro}{Introduction}

Machine learning has made enormous strides in recent years on multiple fronts. Of course, there are deep  neural networks (DNNs) for visual recognition~\cite{alexnet,vggnet,resnet, inception,seg1,seg2,Goodfellow2014,zhu2017unpaired,isola2017image, wassersteingan2017}.
Separately, \emph{gradient boosting (GB)}~\cite{Friedman2001} with decision trees has continued to be a dominant method for table-based datasets (e.g., relational databases and spreadsheets which are indispensable in enterprises for capturing sales, revenue, payroll, inventory, and insurance data), especially those with categorical or discrete data as is common in many analytics (e.g., recommendation systems, insurance, regression). 
The advances and open-source software for offline training of GB models
(e.g., \cite{xgboost, lightgbm, catboost, ThunderGBM}) have resulted in the 
models' extensive use. For instance,  XGBoost~\cite{xgboost}, which 
has over 5800 citations, won the prestigious Infoworld's 2019 Technology of the Year award, the first place in the ACM 2017 RecSys Challenge (for recommendation systems),  and numerous Kaggle awards over the years, as stated in the 
`Awesome XGBoost' Web site \cite{awesome}, as well as the 2016 John Chambers award~\cite{xgboost-wikipedia}. XGBoost alone is used by over 3200 companies, including eBay, Capital One, and  Nvidia~\cite{hgdata}.
Facebook~\cite{facebook-adclick} and  Bing~\cite{bing-adclick} have
used GB to predict advertisement clicks, a key revenue source.
In Kaggle 2020 public solutions, GB is used in 4 out of 18 winners as opposed to
neural networks (NNs) in 11. While numerous accelerator proposals
(e.g., \cite{TPU,diannao,eyeriss,cnvlutin,fusion,scnn,sparten} \hl) exist for 
NNs, the most popular ML method, only a few exist for GB, the second-most popular 
method (\secref{prior-FPGA}). Anecdotally, a google search of GB reveals its indisputable importance for
tabular data.  In addition to its empirical success, GB's  rich
theoretical foundation  gives confidence in the method \cite{Schapire2012}.

The core idea of boosting is to combine many simple or \emph{weak} models (i.e. low-accuracy models) to form a \emph{strong} model (i.e., a highly-accurate model).
Unlike DNNs, where the parameters of the model are trained via gradient descent, GB~\cite{Friedman2001} refines an aggregate function (the {\em strong} model) by 
 taking approximate gradient steps in the function space (by incrementally adding  {\em weak} models).

Traditionally, the computational bottleneck has been the determination of the next best decision point to grow  the current tree.
In the exact tree algorithm, all possible decision points along all possible fields must be considered, which naively costs $O(nd)$ where $n$ is the number of instances and $d$ is the number of fields (assuming the field values are already sorted).
A  common method to avoid this high computational cost is by replacing the fields with discretized values via histograms that approximate the points---effectively reducing the number of possible decision points from $n$ to the number of histogram bins $k$, where $k \ll n$ \cite{xgboost, fastbdt, lightgbm}.
These histogram-based methods have been shown empirically to be much faster with only a slight loss of accuracy~\cite{lightgbm, fastbdt}.

While training GB models is time-consuming (e.g., a few to several hours on datasets with  hundreds of millions to billions of records), a {\em single} input record inference is fairly lightweight (e.g.,  a few microseconds on a multicore) amounting to traversing a few hundred shallow regression trees (e.g., 500, 6-deep trees). Further, increasing the accuracy typically requires more training data which in turn requires larger models (more trees) to avoid over-fitting. Both more training data and larger models lead to longer training times. 
As such, we propose an accelerator for GB  targeting, for instance,  (a) datacenters that offer training as a service, and (b) batch inference (e.g., offline analytics).    {\em Specifically} targeting DNN training, the TPU V2 and V3 extend the TPU  by adding floating-point support and a high-bandwidth network, and Nvidia's DGX includes a high-bandwidth switch (NVSwitch). Thus, providing hardware support {\em solely} for time-consuming training, though offline, is common. 
The GB training algorithm grows the model one decision tree at a time; and grows each tree by starting with its root and splitting nodes to form a tree. The training time for GB models 
is dominated by three key steps -- histogram binning of records' gradient statistics, single-predicate evaluation, and evaluating a record on a single tree. \secref{background} expands  on the training algorithm, including the above listed steps. Each of these steps  operates (repeatedly) on millions of records in the training set. As such, we target these steps for acceleration.
There is one other step -- evaluating histogram bins to identify a split point -- that is algorithmically significant for training, but operates on 
smaller datastructures (e.g., 1000s of histogram bins, instead of millions of records) and hence, accounts for a smaller fraction of total training time (e.g., 2\%-10\%).
Because this step is short, and further because the step involves formulae that are subject to change (based on loss function, for example), we offload it to the host CPU.

The training algorithm has enormous parallelism,  across both the {\em fields} of an input record (e.g., 100 {\em fields} expanded to some 4000 {\em features} due to one-hot encoding of {\em categorical data}, as explained in~\secref{background})  and different input records (e.g., 200 million).  However, an input record may fall into different histogram bins for  different fields (e.g., bin 20 for field 1 and bin 38 for field 2) and 
the input records may fall into different histogram bins for a field, 
imposing significant irregularity in histogram access. On the positive side, the data structures (e.g., histograms and decision trees) are small enough to fit on-chip (e.g., under 2 MB). 
While all the records are binned at the root, the other vertices receive
only a subset of the records, filtered by the predicates in the path from the root to a given vertex. This subsetting makes the input record accesses in the binning and single-predicate evaluation irregular. 
Further, the single-predicate  evaluation and the one-tree traversal use only one field and a subset of fields of a record, respectively, potentially wasting memory bandwidth on the unused fields.  Similarly, batch inference has parallelism across the decision trees traversed by each record and across different records.  Batch inference is also irregular in that the records may take different paths through a tree and, of course, different trees' predicates are different from each other.
Thus, irrespective of training or batch inference, GB has {\em enormous fine-grain, irregular access parallelism} where the compute work is mostly accessing histogram bins or decision tree vertices. This data access parallelism is in sharp contrast to the multiply-accumulate parallelism of DNNs~\cite{TPU}. 
 
While CPUs can handle irregularity, their  parallelism is limited (e.g., 64 threads in a 32-core multicore). GPUs, on the other hand, can exploit immense parallelism but face difficulty with irregularity. While the GPU Shared Memory can accommodate some irregularity, its limited capacity amounts to either almost entirely turning off multithreading or incurring considerable underutilization (e.g., over 90\% idle GPU lanes), as we discuss in~\secref{background}. Clearly, DNN accelerators, such as the TPU, are a mismatch for GB.

Based on the above workload characteristics, we propose an accelerator for GB, called \name, which makes the following contributions: 

\begin{itemize}

\item To match the {\em fine-grained, irregular parallelism in accessing many small data structures}, we propose a {\em scalable sea-of-small-SRAMs} architecture. The SRAMs hold the histograms (for training) or the decision trees (for training or for batch inference), as appropriate. 
Each SRAM has an associated floating-point unit to perform some computation. 
The number of SRAMs determines the available degree of parallelism required to match the memory bandwidth --  across the fields of a record and across the input records (e.g., scale to 3200 SRAMs processing 64 fields per record and 50 records in parallel). 

\item Naively mapping one-hot encoded features to SRAMs (by packing the histogram bins into SRAMs, for example) would result in idle SRAMs as a given record would have only one active feature out of all the one-hot encoded features corresponding to a categorical field. 
Further, for accurate binning, the training algorithm bins the records with missing fields in a default histogram bin for each such field. Thus, while the one-hot encoded features appear to be sparse, the  higher-level fields are dense where every record has every field exactly once.  We exploit this observation to map a high-level field -- i.e., all the corresponding one-hot encoded features including the bin for absence --  to an SRAM,  achieving full SRAM  bandwidth utilization (exactly one access per SRAM). Though dense and parallel, the accesses remain irregular, rendering GPUs ineffective and CPUs insufficient.

\item Finally, to save memory bandwidth in single-predicate evaluation and one-tree traversal in training, we employ {\em a per-field column-major format} for the input records, in addition to the natural {\em per-record row-major format}.  Column-major format is well-known, our novelty is the redundancy. The input records already undergo offline pre-processing in software. The {\em redundant} format adds a little to the pre-processing time which is amortized over as many scans of the input data as the trees built by training. 

\end{itemize}

We evaluate \name via simulations, and an ASIC synthesis. 
Our simulations reveal that 
\name achieves 11.4x speedup and 6.4x speedup over an ideal 32-core multicore and an ideal GPU, respectively. Based on ASIC synthesis of FPGA-validated RTL using 45 nm technology, we estimate a \name chip to occupy 60 mm$^2$ of area and  dissipate 23 W of power when operating at  1-GHz clock speed. 

\putsec{background}{Gradient Boosting Trees}

Our description is based on state-of-the-art implementations of gradient boosting (GB) (e.g., ~\cite{xgboost,lightgbm,catboost,xgboost}). 
We summarize GB with a focus on (1) offering high-level intuition on its operations, and (2) identifying key workload behavior that is relevant to \name's design; we do not aim to describe {\em all} of GB's details.

\putsubsec{xgboost-training}{GB: Model and Training}

The tree ensemble models used by GB  have the following characteristics:
\begin{itemize}
   
    \item The ensemble has $K$ trees with decision rules in the interior nodes and weights ($w$) associated with the leaves.
     \item The model operates on table-based datasets. Each row of the table corresponds to a record and each column corresponds to a field.
    \item There are $n$ training records, with known ``golden'' outputs; The $i^{th}$ record has a known output value of $y_i$.
    \item Prediction/inference involves passing a record through all the individual trees to generate a weak prediction per tree (the $w$ associated with the leaf) which can then be combined for the final prediction ($\hat{y}_i$), which is the strong prediction. \figref{treeexample} shows a simple tree ensemble model with $K=2$ trees. It illustrates the computation of the strong prediction for the red and blue customer records.
    \item GB is agnostic about the loss function as long as the loss function $l(\hat{y}_i, y_i)$ is differentiable and convex. In addition to the loss function, the objective function (that GB minimizes during training) includes terms for penalizing tree complexity and for weight regularization.
\end{itemize}
 
\figput[\linewidth]{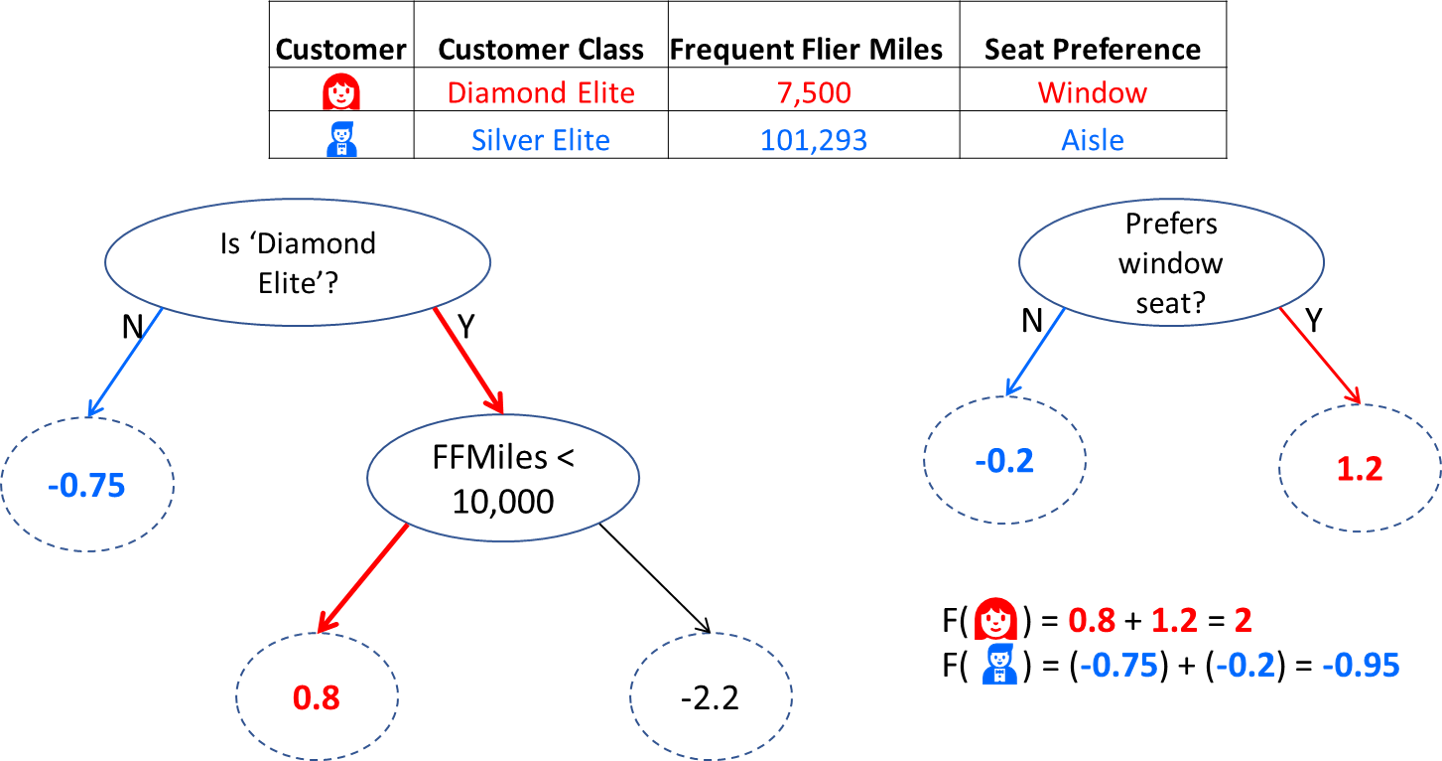}{}{Tree Ensemble Model: Prediction example}

The GB training algorithm arrives at the final $K$-tree model by incrementally growing the ensemble one tree at a time. Each
tree is in turn grown by splitting leaf vertices, one leaf at a time as shown in~\tabref{xgboost-algo}. 
At a high level, GB training determines the split points of a leaf  based on the distribution of first- and second-order gradient statistics ($g_i$ and $h_i$) of the training records. 
We expand on this process to identify the key computational tasks. 

We use the example shown in \figref{groupbyfield} which defines the 
fields of a fictional airline's frequent-flier database; the same database used earlier in \figref{treeexample}. (Only the field
definitions are shown here; the data is not shown in \figref{groupbyfield}.) 
The fields include categorical data (fields 1 and 2
 shown in blue and amber, respectively, in \figref{groupbyfield}) as well as numerical fields (field 3 in green). 
For simplicity, we first show a logical view of the computation. The real implementation optimizes and eliminates some of the operations where possible.
In preparation for training, the input records are (pre-)processed  in software  (1) to discretize  %and one-hot encode 
floating-point fields into some number of bins (e.g., 256 bins, including one bin for records with a missing field), (2)  to one-hot encode  categorical fields  (i.e., the 3-category field 1 is expanded into three mutually-exclusive `yes-no' binary {\em features}, where each binary feature has two bins), and (3) to include an `absent' bin for each categorical field; not all records have values in all the categorical fields. 
For high inference accuracy, the algorithm  accounts for missing features using the absent bins.
Naive one-hot encoding increases the size of each record resulting in much higher memory bandwidth demand (e.g., 40 original fields expand to 200 features). However, a key pre-processing optimization  ensures that only one bin per field sees a gradient statistic update by effectively counting only the `yes' features and reconstructing the `no' features  by subtracting from the overall gradient statistic summation~\cite{lightgbm}. Similarly, the 'absent' bin can also be elided. 
We incorporate this optimization into our GB baseline.

\tabput{xgboost-algo}{\begin{tabular}{|c|p{2.75in}|}
\hline
\bf Step & \bf Details \\
\hline
\rawstep{1} & {\bf histogram-binning} of the gradient statistics of each field of the input records that reach the leaf  under consideration (by default, all records reach the root) \\ \hline
\rawstep{2} & based on the gradient statistic summations in each histogram bin, {\bf finding the best bin to split a leaf}; a split point adds children to the current leaf using a predicate based on a feature and a split criterion (e.g., age $\ge$ 22) \\ \hline
\rawstep{3}& {\bf applying the newly-determined splitting predicate} to partition the records reaching the new vertex into ``predicate true'' and ``predicate false'' subsets for the next iteration of the leaf-splitting algorithm \\ \hline
\rawstep{4} & repeating the first three steps until the tree reaches the desired depth (or until the loss does not improve) \\ \hline
\rawstep{5} & {\bf evaluating the completed tree} to determine the residual loss of the model with the newly added tree, and to compute gradient statistics for the updated model \\  \hline
\rawstep{6} & repeating the above steps for a new tree if the loss continues to decrease \\
\hline
\end{tabular}}{GB steps {\bf (key data structure in each step)}}

% The pre-processing reduces
The histogram-binning of the relevant records  in \step{1} (\tabref{xgboost-algo}) add a given record's gradient statistics ($g_i$ and $h_i$) to the bin's summation ($G$ and $H$) respectively. \step{2}
identifies the feature and  bin within the feature  that should be used as the split point for the children  of the current leaf. To do so, this step evaluates every bin of every feature as a potential split point. For a given feature, the algorithm considers the bins in left to right order and
starts with the split point to the left of all the bins 
so  that  the left cumulative count  is  zero and the right cumulative count is  all the records reaching the leaf. The algorithm then successively moves the split point right by one bin and  adds that bin's count, $G$ and $H$ to the left cumulative bucket and subtracts the bin's quantities from right cumulative bucket. 
\figref{evalsplit} illustrates the summation of the gradient statistics 
to the left and right of the split-point under consideration -- the upper bin boundary of the $i^{th}$ bin. (Though not shown, GB considers 
placing records with missing fields in both the left and the right sub-trees to pick the best split option.)
The quality of this split is 
evaluated based on the left and right cumulative buckets. 
The algorithm greedily picks the best split point across all fields and split points.  
We note that the function to evaluate the quality of a split is often complex (i.e., hardware-unfriendly) and may vary across implementations. The algorithm may choose not to create any children for the current leaf if the loss improvement is exceeded by the complexity cost associated with the children (larger trees have higher complexity cost).

\step{3} in~\tabref{xgboost-algo} applies the newly-chosen predicate (corresponding to the split-point) to partition the relevant records (i.e., those that reach the current leaf after being filtered by the predicates in the path from the root to the leaf). The records are partitioned into predicate-true and predicate-false subsets which are used later as the relevant records for exploring the next vertices on the corresponding paths. The subsets are stored in memory. An optimization in \step{1} is that the histogram-binning can be performed only  for the child vertex with the smaller subset and the other child vertex's (with the larger  subset) bin counts, bin $G$ and $H$ can be
calculated by simply subtracting the first child's bins from the parent vertex's bins without any explicit binning at the other child~\cite{xgboost}.
\step{4} repeats steps \rawstep{1} through \rawstep{3} to the relevant records to complete the tree. With a fully-grown tree, \step{5} passes all the input records through the tree to update each record's first and second order gradient statistics ($g_i$ and $h_i$) based on the new tree's prediction
compared to the ground truth (i.e., the per-record loss), the loss function, and
the record's  previous $g_i$ and $h_i$.  
This step also updates the new total loss across all the records for  the new and old trees put together. 

\figput{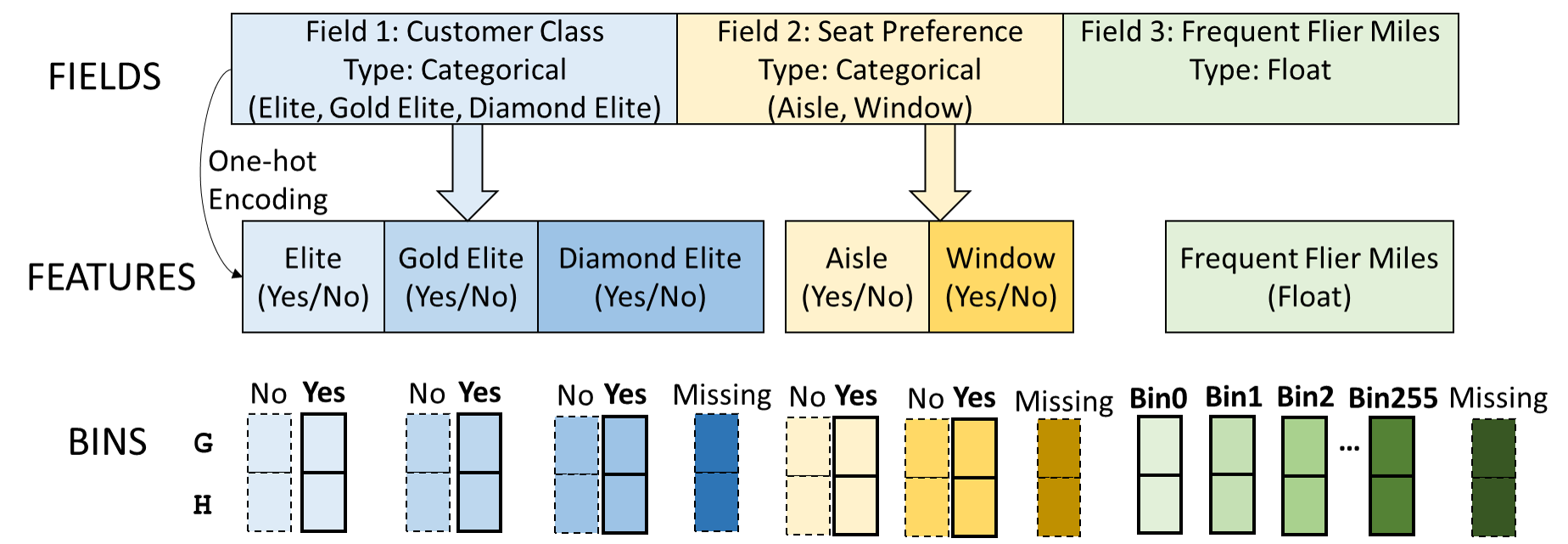}{}{Fields, Features and Histogram bins}

GB implementations can  be configured to proceed vertex by vertex or level by level (i.e.,  explore together all the valid vertices at a level though some vertices may be missing). The above
assumes the former, whereas the latter streams in all the input records and histogram-bins the relevant records at each vertex. Because multiple vertices are explored together, this configuration maintains a separate histogram per vertex. 

In addition to the above qualitative description, we quantify the fraction of time spent in the algorithm's steps  later in \secref{method}  and \secref{breakdown}.

\putsubsec{inference}{Batch Inference}

In batch inference, each record traverses all the decision trees. In a tree, each vertex's predicate is evaluated for the record to determine the next vertex. At the leaf, the tree outputs a value.
All the trees' outputs are combined to compute the final prediction for the record.
Unlike training, inference involves only the tree traversal. 

\putsubsec{workload}{Parallelism and Access Patterns}
There is enormous fine-grain parallelism in training both across the features of a record (e.g., 10s-1000s) and across
different records (e.g., hundremillions). As mentioned in~\secref{intro}, the computation and its parallelism are almost entirely data-access heavy as opposed to computation-heavy like DNNs. 
\step{1} in~\tabref{xgboost-algo} has both intra- and inter-record parallelism. The histograms across the features can be updated in parallel for {\em intra-record} parallelism. Fortunately, the histograms can fit on-chip (e.g., 2-8 MB).
However, because different features of a record often update different bins (e..g, feature 1 may update bin 8, feature 2 may update bin 0), the accesses are both fine-grained and irregular.   An easy way to exploit {\em inter-record} parallelism is to replicate  the histograms followed by a reduction across each histogram's replicas. 
\step{2} is short  as it iterates over all the features' bins which can be processed in parallel but  are far fewer than the input records. 
\step{3}'s parallelism is inter-record where the new predicate is evaluated for each record.   
\step{5}'s parallelism is also inter-record where the new tree is traversed for each record. 
However, because only subsets of the records are likely relevant to vertices other the root, the records' access is irregular (non-contiguous) in steps \rawstep{1}, \rawstep{3}, and \rawstep{5}. Further,  \step{5}'s tree traversal is also highly irregular because different records would follow different paths through the tree. While steps \rawstep{1} and \rawstep{3} do not have any load imbalance issues, different records in \step{5}  may see slightly different path lengths in some cases. Because the training algorithm grows the trees up to a pre-specified depth, the path lengths are mostly similar (but not  the same because some paths may be terminated early due to non-profitability).
Fortunately, such modest load imbalance can be handled by  averaging out over multiple records. 
Finally, steps \rawstep{3} and \rawstep{5} use only a feature and a subset of the features of the records, respectively, potentially wasting memory bandwidth. 

In batch inference, there are both abundant inter-tree parallelism (e.g., 500-1000) and inter-record parallelism (e.g., hundreds of thousands in small batches  to millions in large batches). However, each record takes a different path through the trees implying significant fine-grained but irregular access parallelism, like training. Further, the path lengths through different trees may be slightly different for different records. Luckily, the load can be balanced by averaging across the records of the batch, like \step{5} in training. 

\figput{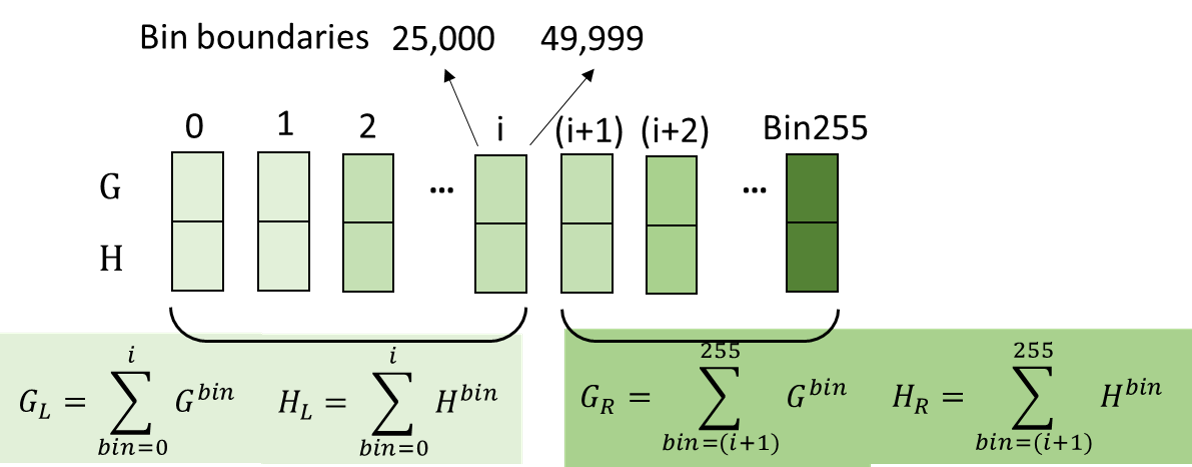}{}{Evaluating a split for {\tt(ffmiles >=50,000)}}

\putsubsechl{cpu-gpu}{Why are multicores and GPUs insufficient?}

While there are fast implementations of GB training on  multicores and GPUs, these implementations leave significant opportunity unrealized. The multicore implementation of GB exploits parallelism and cache locality in training while dealing with the irregularity. The input records are partitioned among the threads each of which has a private version of the histograms of \step{1} (\tabref{xgboost-algo}) at the end of which the histograms are reduced.  \step{3} is  parallelized by partitioning the input records and replicating the current tree among the threads. The independent threads can handle the  irregular accesses to  the histograms (\step{1}), the tree (\step{5}),  and the records (steps \rawstep{1}, \rawstep{3}, and \rawstep{5}). However, the multicore's modest  parallelism (e.g., 32 threads) and limited on-chip cache to hold the replicated histograms means multicores exploit only a small fraction of the parallelism available in the application (100-500 intra-record and abundant inter-record parallelism). 

Similarly, in batch inference,  multicores can exploit only a fraction of the huge inter-tree and inter-record parallelism. 

In contrast, GPUs can exploit abundant fine-grained parallelism in training as long as the parallelism is regular.
However, the data-dependent histogram bin updates are irregular. GPUs can even handle memory-access irregularity as long as the footprint of the irregular accesses fits in 
special high-bandwidth structures like the GPUs' Shared Memory, which is typically limited to 48KB-96KB. 
Unfortunately, even that option is not usable for GB training because of the read-modify-write nature of histogram bin updates. Specifically, multi-threading, which is fundamentally required for GPUs to achieve high performance by hiding memory latency, interacts poorly with read-modify-write updates. The interleaving of GPU warps due to multi-threading will lead to incorrect results. Each of the two possible ways to overcome this read-modify-write problem results in performance degradation in their own way. The first way  is to use locks to maintain atomicity of the read-modify-write components of bin updates. Locking is expensive as it effectively limits multithreading by disallowing fine-grained thread interleaving. The second way is to privatize the 
bin state so that multithreading updates are going to different copies. However, privatization leads to a linear (in the degree of multithreading)
increase in the number of copies of the bin state, which will not fit in GPUs' Shared Memory. For example, each warp will need to maintain a private copy of the histogram bin counters. Even  the smallest of our benchmarks (\higgs) has 28 numerical features yielding 7K bins (using 256 bins/field) of 8 bytes each -- i.e., 56 KB per warp. The typical per-SM Shared Memory under 96 KB can accommodate at most one warp  (i.e., no multi-threading) which would eliminate the GPU's ability to tolerate latencies. Alternatively, the GPUs would need an impractically-large 1.75 MB of Shared Memory to accommodate 32 warps.

In summary, it is difficult to achieve high performance for GB training % (which uses data-dependent, irregular, read-modify-write accesses) 
on GPUs because of the above multiple constraints of GPUs.
% (need for multi-threading, need for regular accesses, and limited Shared Memory capacity). 
Indeed,  GB implementations on GPUs achieve modest (around 2x)  speedups over the multicore implementations, which we validate in \secref{sanity}. 
For example, a recent GPU-optimized GB implementation achieves around 
2.3x mean speedup over a 20-core multicore implementation~\cite{ThunderGBM}.

For batch inference on GPUs, multithreading is possible because there are no read-modify-write problems. However, SIMT parallelism is significantly degraded as different trees (and different records) traverse different tree paths. The lack of SIMT parallelism can result in poor GPU performance. Using Shared Memory to accommodate irregular accesses requires holding as many trees (or copies) as the threads which will not fit. 

\putsubsechl{prior-FPGA}{Previous FPGA-based GB acceleration}

One FPGA-based work~\cite{GB-Tanaka} parallelizes only across records like a multicore by holding histogram copies which degrade on-chip capacity. Booster's intra-record parallelism is higher than that of this approach to which we compare in~\secref{perf}.
Another work~\cite{GB-Sadasue}  is just a one-page paper which pipelines the processing but includes few other details.

\putsec{booster}{\name} 

Recall from~\secref{intro} that \name's key contributions are: (1) a {\em scalable sea of small SRAMs} to match GB trees' abundant, fine-grained, irregular, small-data-structure access parallelism (e.g., scale to 3200 SRAMs),
(2) mapping the gradient-accumulating histogram bins (step \rawstep{1}) to SRAMs using a {\em group-by-field} strategy that outperforms a naive greedy-packing strategy, 
and (3) employing a redundant {\em per-field column-major format} in addition to the natural per-record row-major format to save memory bandwidth in single-predicate evaluation and one-tree traversal (steps \rawstep{3} and \rawstep{5}). 

\putsubsec{mapping}{Group-by-field mapping of bins to SRAMs}

As discussed in \secref{background}, 
the numerical fields are associated with a number of bins. 
\figref{mapping} illustrates the bins for the three fields of the frequent-flier database. 
The two categorical fields have three and two `yes' bins respectively. Recall from \secref{background} that the `no' bin gradient summations can be reconstructed.
For ease of illustration, \figref{mapping} assumes that FP features are discretized to six (6) bins; a real application would typically use 128-256 bins in practice.

\figput{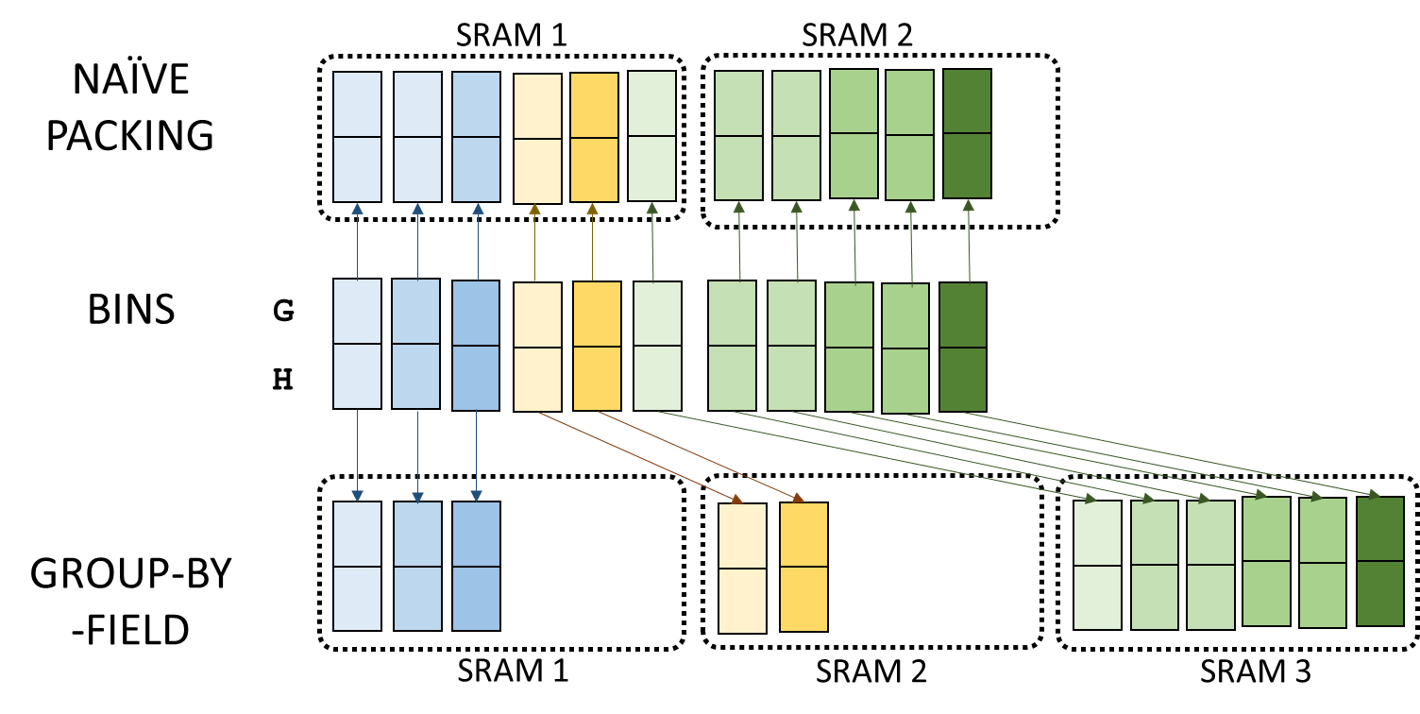}{}{Group-by-field mapping to SRAMs}

The placement of the histogram bins in SRAMs has a significant impact on work serialization and balance in step~\rawstep{1}.
For example, if histogram bins belonging to multiple fields are placed in the same SRAM, multiple updates to those bins have to be serialized. However, histogram bins placed on different SRAMs can be updated concurrently.
\name's sea-of-SRAMs approach helps harness this parallelism.

In addition, the mapping of bins to SRAMs has an impact on SRAM load balance.
{\em Naively packing} the bin state in SRAMs based on capacity results in load imbalance and underutilization.
\figref{mapping} illustrates the naive packing approach assuming an SRAM capacity of six bins. Bins are allocated to SRAMs (shown in dashed boundaries) till the SRAM if filled. 
Under this assignment,  all three gradient updates corresponding to a record may be assigned to the first SRAM as shown in \figref{mapping} whereas the second SRAM may have no updates to process. 

We observe that there is {\em exactly}  one update across all the bins of a field, whether numerical, categorical, or missing (recall from~\secref{background} that each field has a default `absent' bin).
\name uses a group-by-field mapping wherein
all the histogram bins of a field are mapped to a single SRAM, utilizing near 100\% SRAM and DRAM bandwidths. \figref{mapping} shows that all bins that are shades of blue (or amber or green) which are associated with a single field are mapped to one SRAM. (If the number of bins associated with a single field exceed the capacity of an SRAM bank, those bins may be spread across multiple SRAMs (details in~\secref{extension}). Our goal is to avoid bins of multiple fields sharing an SRAM bank as that leads to (avoidable) serialization.

\putsubsec{microarch}{\name microarchitecture}

\name's microarchitecture consists of  a sea of small SRAMs each of which has a floating-point adder for updating $G$ and $H$ (\figref{boosterarch}). Each {\em Booster Unit (BU)} comprises one SRAM and the associated adder. Multiple BUs are organized in a cluster (e.g., 64)  which may be replicated (e.g., 8).

For the current tree leaf, the single-predicate step in the {\em previous} iteration of the inner step \rawstep{4} loop produces a stream of pointers to the relevant records (all the records are relevant for the root).
For the histogram-binning step (step \rawstep{1}),
\name fetches each relevant input record  (one or more memory blocks, say 64 bytes) using the pointers, distributes each field to a BU, and broadcasts the record's $g_i$ and $h_i$ to each BU. The broadcast bus in~\figref{boosterarch} performs a {\em logical} broadcast implemented as a simple, pipelined broadcast over point-to-point links (e.g., 16 BUs per link).
Because the input has millions of  records, this pipeline's fill and drain overheads are negligible (e.g., 3200/16 = 200 cycles).  
The distribution of the fields occurs in a simple, fixed left-to-right order of the fields and SRAMs. 
The fixed order implies that simple one-to-one buses from the  fetch buffer to the BUs suffice. The record format also specifies the starting feature number per field so that each BU can subtract this number from a record's feature number for the field to locate the feature in the SRAM.
\name employs simple double-buffering to hide completely the fetch latency. Because the full set of pointers to the records relevant for every step is known a priori, the implicit prefetch of double-buffering removes memory latency as an issue.  Further, though the relevant records would likely not be contiguous in memory, each record is one or more memory blocks of contiguous bytes, thus achieving good memory bandwidth. 

For step \rawstep{1}), each SRAM holds a field's histogram bin counters, $G$ and $H$ for each bin (\tabref{arch-data}). Each BU simply increments its incoming bin's counter, and adds the record's $G$ and $H$ to the bin's $G$ and $H$. For more parallelism, the records can be partitioned among the clusters so that each cluster generates a set of histograms which are reduced at the end of the step.  Because choosing the best split point (\step{2}) (a) is light-weight whose execution time is proportional only to the number of bins (thousands) and not the number of records (millions), (b) is often complex (i.e., hardware-unfriendly), and (c) may vary across implementations (as noted in~\secref{xgboost-training}),  we offload the step along with the reduction at \step{1} end to the host.  

\figput{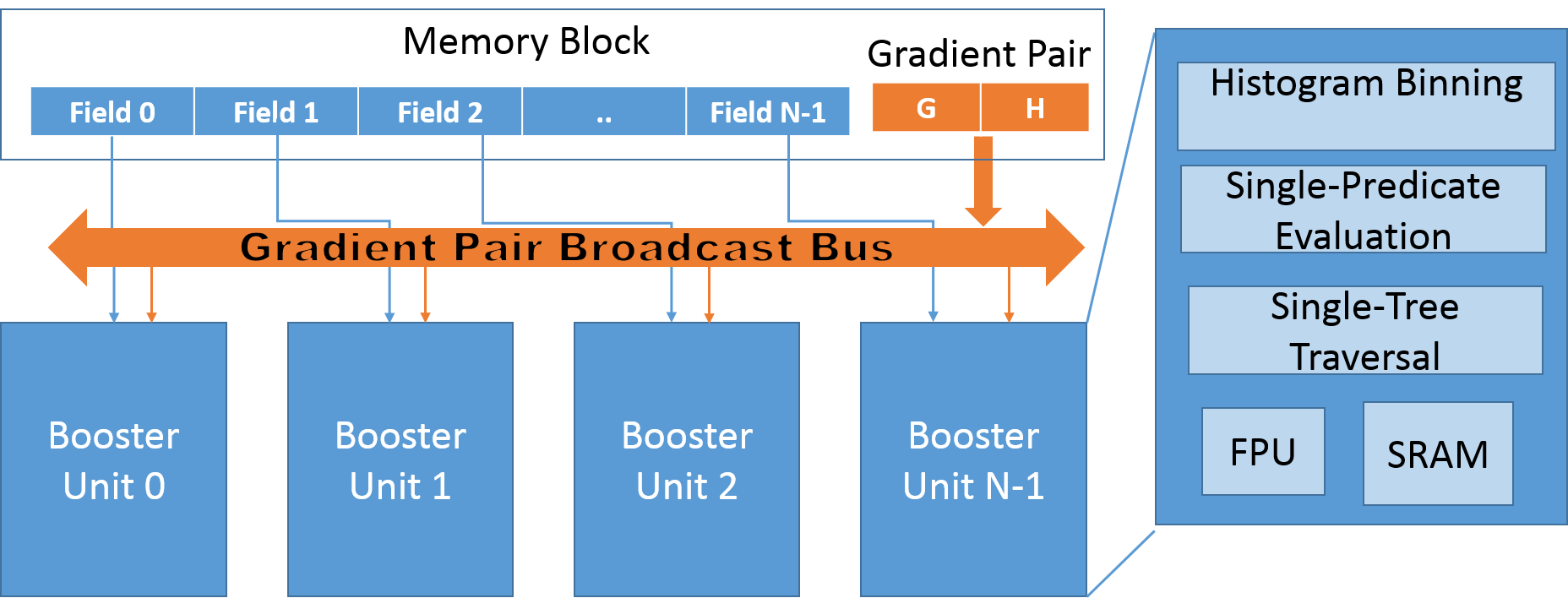}{}{\name microarchitecture}

The best split point is expressed as a predicate used in  the single-predicate step (\step{3}). In this step, \name first receives the predicate as a field number, a bin number, and a condition (e.g., {\tt ffmiles}  $\ge$ {\tt 50,000}) may
be encoded as {\tt (Field-3 $\ge $ upper-bin-boundary($Bin_i$))}, as shown in \figref{groupbyfield}. This predicate is broadcast (replicated) to all the BUs of one or more clusters (\tabref{arch-data}). The field number enables \name to fetch only the relevant field of each relevant record in the redundant {\em per-field column-major format}.  As before, the relevant single-field columns are distributed to the BUs which evaluate the predicate for each record and place the pointer in either the ``predicate true'' or the ``predicate false'' buffers.  The BUs' buffers are streamed out to off-chip memory. 
Like the input records, the output streams are also double-buffered. Unlike the full relevant records, the relevant single-field  columns would likely be more non-contiguous (e.g., in a memory block of a single-field column, only a subset may be relevant). Nevertheless, our redundant format saves significant off-chip memory bandwidth compared to the original per-record row-major format which would fetch the whole record but use only one field. 
\step{4}  iterates over steps \rawstep{1} through \rawstep{3} to grow the tree to the desired depth (or stop if the loss does not improve).

For \step{5} (one-tree traversal),  we apply the well-known idea of mapping the newly-grown tree to a table where each entry captures a vertex by encoding its predicate (like the single-predicate step with a slight modification) and pointers to the vertex's left and right children. The table is broadcast to all the BUs and replicated in the SRAMs (\tabref{arch-data}). Based on the set of fields relevant for the tree's predicates, \name fetches the corresponding single-field columns, and {\em G} and {\em H} of all the records.  The modification in the predicates is that instead of using the original field numbers, the predicates use a renumbering among the relevant fields
(e.g., the original field 228 may be renumbered as the new field 7 among the fields relevant to the tree). Each BU operates on a record which traverses the tree sequentially by looking up the table in the SRAM. The new field number in the predicate indicates the field to be looked up in the record. As the record exits the tree, the record's prediction is compared with the ground truth to update the record's  {\em G} and {\em H} and the total loss. 
The records' new {\em G} and {\em H} are double-buffered and  written back to off-chip memory in a stream. This stream efficiency motivates storing these fields separately.

\begin{table*}
\parbox{2in}{
\caption{\label{tab:arch-data}Data structure to BU mapping}
\begin{tabular}{|c|p{1.4 in}|}
\hline
\bf Step & \bf Data structure mapping \\
\hline
\rawstep{1} & Map one field's bins to one BU \\ \hline
\rawstep{2} & Offload \\ \hline
\rawstep{3} & Replicate predicate at each BU \\ \hline
\rawstep{4} & Repeat steps \rawstep{1}-\rawstep{3}\\ \hline
\rawstep{5} & Replicate tree at each BU \\  \hline
\rawstep{6} & Repeat steps \rawstep{1}-\rawstep{5}\\
\hline
\end{tabular}
}
\parbox{4in}{
\caption{\label{tab:prelim-dataset}Dataset and Model Characteristics}
\begin{tabular}{|c|c|c|c||c|c|p{1.25in}|}
\hline
\bf Name & \bf \#Records & \multicolumn{2}{c|}{\bf \#Fields} & \bf \#Features &  \bf Seq. Time &\bf Comment\\
 & \bf (millions) & \bf All & \bf Categ. & \bf (one-hot) & \bf (mins) &\\
\hline
\iot~\cite{IOT-data} & 7 & 115& 0 & 115  & 15 & Botnet attack detection\\
\higgs~\cite{Higgs} & 10 & 28 & 0 & 28  & 18.5 & Exotic particle collider data\\
\allstate~\cite{Allstate-data}  & 10 & 32& 16 & 4232 & 1.6  & Insurance claim prediction\\
\mq~\cite{Mq2008} & 1 & 46 &0 & 46  & 2.5 & Supervised Ranking\\
\flight~\cite{Flight-data,airline-data} & 10 & 8& 7 & 666 & 5.5 & Flight delay prediction\\
\hline
\end{tabular}
}
\end{table*}

\name's implementation should choose the SRAM size to be the smallest that accommodates all the features of a field, for typical fields. 
The amount of parallelism -- i.e., the number of BUs -- should be enough so that the on-chip work is rate-matched with the off-chip memory bandwidth. 
For example, assuming  400 GB/s memory bandwidth (sustained average bandwidth measured in \secref{method}), 1-GHz clock for \name, and 64-byte blocks, leads to 6.25 blocks supplied every cycle. Each field consumes a byte to specify its feature (i.e., 64 fields in a block). 
 The compute work for every field at its BU is a short integer subtract (to calculate the bin),  an SRAM read, two pipelined floating-point adds,  and an SRAM write, which can fit in, say,  8 cycles.
The total SRAM occupancy associated with $6.25 (blocks) * 64 (fields/block) * 8 (cycles/field) = 3200 (cycles)$. Thus, 3200 SRAMs is adequate to saturate the memory bandwidth. 
\name's high parallelism effectively ensures that it is memory bound with all computation hidden under memory latency (of record reads).

Finally,  FPGAs can support Booster realizations while keeping in mind the usual FPGA-versus-ASIC trade-offs of time to market (FPGAs are better, especially with Cloud-based FPGA resources), power/performance penalties (FPGAs are worse), and volume-dependent costs (FPGAs are less expensive at small volumes). Our contribution is the architecture, {\em independent of ASIC or FPGA implementation} though we show an example ASIC implementation so that we can compare Booster against  high-performance CPUs and GPUs.

\putsubsec{extension}{Microarchitecture extensions}
We consider a few exceptional cases. (1)  If there are more  fields than the SRAMs, then the records are partitioned into subsets of fields with the $G$ and $H$ fetched for every partition. The records are then  histogram-binned (\step{1}) partition by partition 
(a partition of all the records before the next partition). Single-predication evaluation (\step{3}) and  one-tree traversal (\step{5}) are not affected by this condition. (2) The partitioning strategy also applies to cases where the records are larger than the memory block. If the records are smaller than the block then there is some off-chip memory  bandwidth loss, but if the records are smaller than half the block then two records are packed into one block. 
(3) If a field has more features than the SRAM entries, than multiple SRAMs are grouped logically to accommodate the field at a modest underutilization of SRAM bandwidth (each record would update only one of the SRAMs in a group). To maintain the one-to-one mapping between fields and SRAMs, pre-processing   repeats that particular field's bin twice in the record if and only if the field is spread over two SRAMs.  Because each BU subtracts its field's starting feature number from the record's feature number to locate the SRAM entry (\secref{microarch}), the repeated bin  falls within only one of the two BUs’ SRAMs (and outside the other BU's SRAM). 

(4) Conversely, it may not be profitable to pack multiple small fields to the same SRAM to save space because such packing reduces throughput. Each record would have to update the SRAM multiple times while the other SRAMs idle. In cases where the off-chip memory bandwidth limits the overall throughput with some SRAM throughput to spare, such packing may not reduce overall throughput.
Further,  each SRAM must accommodate a floating-point field in case an application has only such fields which typically have 256 features requiring 2 KB, the typical SRAM size. 
Then, such packing would mean larger SRAMs (grouping multiple SRAMs for capacity would reduce overall throughput) which ironically means more on-chip SRAM capacity to achieve higher space efficiency. As such, our results show 89\% capacity utilization and near 100\% SRAM and DRAM bandwidth utilization for our one-field-per-SRAM mapping.  (5) The trees  not fitting in the SRAM  is unlikely given that GB's core idea is to form a strong model by combining many weak models (i.e., shallow trees). While a larger tree can  be partitioned to fit into a group of SRAMs, we leave this case for future work.

\putsubsec{booster-inference}{Batch inference on \name} 
Implementing batch inference on \name is an extension of the one-tree traversal in training with the key difference being multiple decision trees in the former versus only one tree in the latter. In inference, each record traverses all the decision trees each of which is loaded as a  table into a BU. Because each tree's predicates may involve tens of fields of a record (typically different sets of fields for different trees), \name simply broadcasts each full record to all the BUs. At each BU, each record sequentially traverses the tree making as many SRAM accesses as the path length through the tree. Multiple records are double-buffered to hide memory latency. Each tree's output (typically a small value such as -1, 0, or 1) is buffered to be combined with the other trees' outputs for a final prediction. The outputs are also double-buffered so that the next record can start as soon as the previous record exits the tree. Thus, each tree is asynchronous with respect to the others, allowing
load balancing across the trees by averaging over the records. 

Finally, the case of too many trees in batch inference not fitting on-chip can be addressed by distributing the trees to multiple \name chips (in a simple round-robin manner). 

\putsec{method}{Methodology}

\noindent
{\bf Datasets: }
We use the five datasets shown in ~\tabref{prelim-dataset}. For each dataset, we list 
the number of (a) training records,  (b) fields per record, (c) categorical fields, and (d)  features (after one-hot encoding of categorical features). For each dataset, we use sequential runs on an Intel 5$^{th}$ generation CPU instance on Google Cloud to train a model of 500 trees, each with depth of upto 6 (see \tabref{prelim-dataset}). We use sequential runs only to understand the work fractions of the various steps. For our performance comparisons, we use the aggressive \im configuration. 

We use datasets under 10 million records to reduce experiment time (training under 20 minutes). Full-scale datasets are larger. For instance, using Terabyte Click Log \cite{terabyteclicklog}, which  has 1.7 billion records with 13 integer and 26 categorical features,  to train 500 trees exceeds 80 hours on an Intel Xeon \cite{terabyteclicklog-runtime}; and a multi-modal dataset for wearables has 63 million records \cite{WESAD}.  
Unfortunately, large, commercial datasets are often not  publicly available 
(e.g.,  Facebook researchers \cite{facebook-adclick}  state that the data 
is confidential and that ``a small fraction of a day\'s worth of data 
can have many hundreds of millions of instances.''). 
\secref{scaleup} scales up our datasets.

\noindent
{\bf Dominant components of execution time: }
\figref{breakdown} shows a breakdown of the normalized sequential training time corresponding to the steps in XGBoost (\tabref{xgboost-algo}).
We see that \step{1} (histogram binning), \step{3} (single-predicate evaluation), and \step{5} (one-tree traversal) constitute over 98\% of run time except for \mq due to its small dataset. 
In contrast, \step{2} (choosing the split point of a tree vertex)
is short enough to be offloaded to the host. As such, \name's omission of \step{2} from 
acceleration is justified.

We observed an interesting behavior in the two benchmarks with categorical data (\allstate and \flight) that reduces the work for histogram binning (\step{1}). Due to one-hot encoding and the nature of the dataset, the left versus right sub-tree split of the records at a vertex is extremely lopsided (99\%-1\%, for example). In such cases, the histogram binning work is reduced drastically  by binning only the smaller sub-tree and setting the larger sub-tree bins to be  the parent's bins  minus  the smaller sub-tree's bins (\secref{xgboost-training}).
This effect is that \step{1}'s share is smaller for \allstate and \flight in \figref{breakdown} despite their large datasets. 

All benchmarks other than \iot yielded trees with leaves mostly at a depth of 6.  \iot had many shallow trees which increases the relative fraction of \step{1} (see 
\figref{opportunity}) because \step{1} processes larger fractions of records closer to the root. This exceptional behavior of \iot impacts both its training 
and  batch-inference performance, as we show later. 

\figput{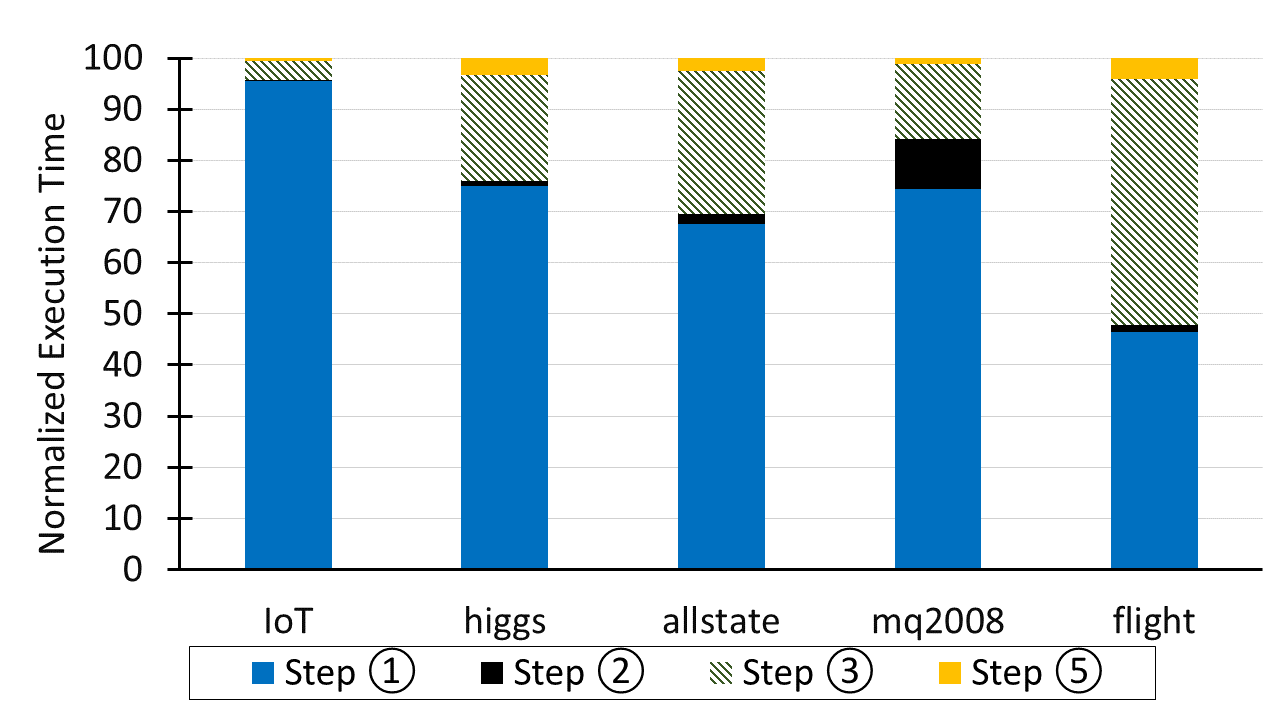}{}{XGBoost sequential execution time breakdown}

\noindent
{\bf Software modifications: }
We implemented and verified that the software changes to XGBoost's input data format (to support the redundant  per-feature column representation) do not affect the numerical results of XGBoost training.

\noindent
{\bf Simulator: }
We build a cycle-level performance simulator for \name. 
Although we do model the delays of our histogram-binning, single-predicate-evaluation, and one-tree traversal (steps \rawstep{1}, \rawstep{3}, and \rawstep{5}) based on our RTL implementation, the delays are hidden entirely under memory latency  by construction (\secref{microarch}). For accurate memory latency and bandwidth simulation, we use DRAMSim2~\cite{dramsim} configured as 
a high-bandwidth, 24-channel memory, whose parameters are derived from the Hynix JESD235 standard and an Nvidia paper ~\cite{gpumem} 
(see \tabref{dramconfig}).  This memory achieves a sustained bandwidth of about 400 GB/s. 

\tabput{dramconfig}{
\begin{tabular}{c|c}
Channels, banks, row     & 24, 16, 1 KB \\
 tCAS-tRP-tRCD-tRAS & 12-12-12-28 \\
\end{tabular}
}{DRAM Configuration}

\noindent
{\bf Simulated systems:} 
Because directly comparing simulated systems and real (multicore or GPU) systems is  not meaningful, 
we simulate two alternatives for a multicore and a GPU called \im and \ig, respectively. We do sanity-check these alternatives against a real multicore and a real GPU.  
For performance comparisons, we conservatively assume 
that the \im and \ig are constrained {\em only} by 32- and 64-way parallelism {\em without any implementation artifacts}
(based on the parallelism limits described in \secref{cpu-gpu}). Because our {\em Ideal} 
configurations assume perfect pipelines and caches, and perfect, convergent SIMT behavior, they are effectively 
upper-bounds on the performance of real multicores and GPUs, respectively.
Both  \im and \ig  use the same memory configuration 
as \name (see \tabref{dramconfig}).  
\secref{sanity} confirms that our \im and \ig configurations indeed outperform a real 32-core multicore (Intel $5^{th}$ generation) and  a real GPU (Nvidia V100), respectively. We also compare to an FPGA-based previous work~\cite{GB-Tanaka}, called {\em Inter-record},  which parallelizes only across records like a multicore (\secref{prior-FPGA}). For fair comparison, we simulate this architecture assuming an ASIC implementation with the same area and clock speed as \name  (i.e., the only difference is the architecture).
Because \name offloads step \rawstep{2} to the host, we add the time for the step on a real 32-core multicore host to  the execution time of all the systems (\name,
\im and \ig). Further, we add the time for the data copy between the host and Booster (to and fro). This
time is of the order of microseconds (PCIe transfers) whereas Booster reduces the total run time from minutes to seconds including the copy. As such, the copying time is negligible.

We model the energy of the \im and \ig configurations to be lower-bounds on energy
consumption (in contrast to upper-bounds on performance). We conservatively model 
(1) the SRAM access energy of each configuration's typical SRAM size (see \tabref{hw-param})
using access activity from our simulator and per-access energy cost from CACTI 7.0~\cite{cacti7.0}. 
Similarly, we model the DRAM access energy based on transfer activity. Such simple models are conservative because
they ignore major energy  overheads in real multicores and GPUs (e.g., multicores' superscalar, out-of-order issue cores, GPUs' large register files and heavily-banked Shared Memory)  that \name does not incur.

\noindent
{\bf RTL implementation, FPGA validation and ASIC Synthesis:} 
We implemented \name in System  Verilog.
We use  Intel's Quartus Prime (v15.0) with Qsys system builder for synthesis. 
We verified the correctness of our implementation using RTL simulation and by running tests on FPGA prototypes using a scaled-down version (8 BUs in  1 cluster). We used an Intel Cyclone IV FPGA with 150K logic elements, which is a modest prototyping FPGA. 
Using our System Verilog implementation, we synthesized an ASIC instance of \name with 1  cluster with 64 BUs using Synopsys's Design Compiler to estimate \name's area, power, and clock speed.
We used the 45-nm technology FreePDK45 PDK/cell library ~\cite{freepdk45} (newer technology nodes are unavailable).  
Because the FreePDK45 does not include a memory compiler, 
we use Cacti 7.0~\cite{cacti7.0} to estimate area and power-per-access for the SRAM tables.

\tabput{hw-param}{
\begin{tabular}{|l|l|l|l|l|l|}
\hline
Configuration& \bf \# cores & Clock  & \multicolumn{2}{|c|}{SRAM}   \\ 
&   &  speed &  size &  energy \\
& & (GHz) & (KB) & (norm.) \\ 
\hline
\hline
Real Multicore$^1$  &  32 cores  &2.2 & 32 KB$^3$ & NA    \\ 
Ideal Multicore  &  32 cores & 2.2 & 32 KB & 1   \\ \hline
Real GPU$^2$  & 80 (64-wide) SMs &  1.5 & 96 KB & NA 	 \\ 
Ideal GPU &  64 (64-wide) SMs  & 2.2  & 96 KB  & 2.64 \\ \hline %48KB & 3.72      \\ \hline
\name & 3200 BUs & 1& 2KB & 0.71 	 \\ \hline

\end{tabular}
}
{Hardware parameters (Google Cloud hardware: $^1$Intel $5^{th}$ generation, $^2$Nvidia V100, $^3$L1D Cache)}

\putsec{results}{Results}
Our results show: (1) performance comparison of
\name, \ig, and \im configurations, (2) execution time breakdown for those schemes to better explain the observed performance trends, (3) the relative performance impacts of the various Booster components, (4) energy comparison,  and  (5) area and power results from ASIC synthesis.

\putsubsec{perf}{Performance}

\figref{main} shows 
the speedup achieved by \ig, the Inter-record scheme (IR)~\cite{GB-Tanaka}  (\secref{prior-FPGA} and \secref{method}), and  \name over the \im baseline (Y-axis) for each of the 
five benchmarks.  
In addition, we also include the geometric mean speedups. 

\ig achieves modest speedups between 1.6x and 1.9x over \im across all benchmarks. We have also validated similar speedups on real hardware (see \secref{sanity}).
IR exploits only inter-record parallelism like multicores, requiring GB histogram copies which are area-inefficient. For the same  area as \name, IR can fit 271 copies/processing units for \higgs and 179 for \mq. While an ASIC cannot have different processing units for different benchmarks, we conservatively assume so as IR's original FPGA-based implementation can achieve this flexibility.
For the other benchmarks, even one copy does not fit, a case not considered by IR which shows results only for \higgs with FPGA-bounded 64 copies. At the same clock speed as \name, IR achieves some modest speedups over \im which essentially has 32 copies.
IR's original speedup for \higgs~\cite{GB-Tanaka} is higher because of (1) holding on-FPGA all 10,000 input records (zero DRAM traffic, infeasible for millions of records), and (2) weaker base cases of 12-thread CPU and Nvidia 1080TI GPU.

Booster achieves speedups varying from 30.6x (for \iot) to 4.6x (for \flight), with a mean speedup of 11.4x, over the aggressive \im baseline. Because \im and \ig are limited only by their exploited parallelism without any implementation artifacts (32- and 64-way, respectively), these speedups isolate the impact of Booster's massive parallelism (3200-way). Similarly, IR's lower parallelism places IR well behind \name.  
For the accelerated steps, \name hits the memory bandwidth limit
by capturing high SRAM parallelism. As such, the overall speedup is effectively limited by a combination of bandwidth limits for the accelerated steps and Amdahl's law limits of  the un-accelerated portion. 

Finally, we note that the performance trends are as expected; we see higher speedups on larger datasets (e.g., \iot, \higgs). Smaller datasets (e.g., \mq) or larger datasets that behave like smaller datasets (e.g., \allstate and \flight)  due to the presence of categorical data and skewed data distributions (see \secref{method}), achieve lower speedups. 
Later, in \secref{scaleup}, we confirm the dependence of speedup on dataset size by scaling up the datasets.

\figput{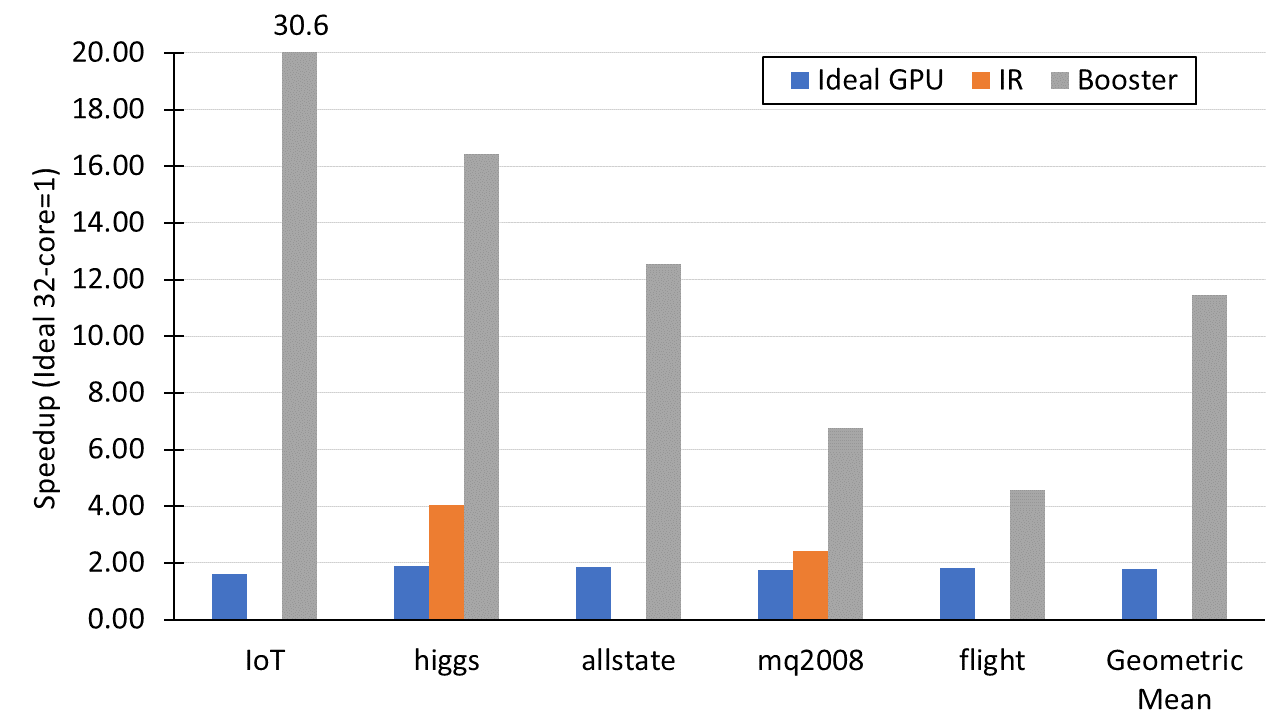}{}{Performance Comparison}

\figput{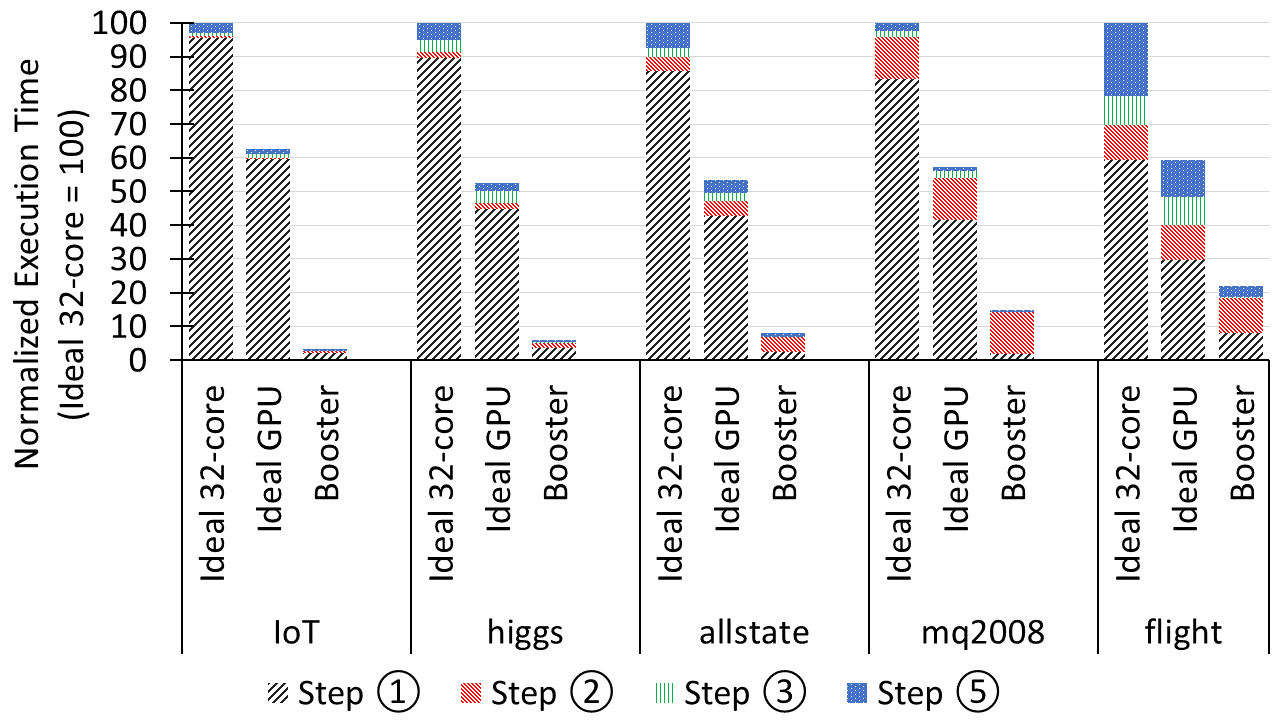}{}{Execution Time Breakdown}

\putsubsec{breakdown}{Execution time breakdown}
%XX OLD TEXT XX
\figref{breakdown} shows the execution time breakdown (Y-axis) of the three architectures normalized to that of  \im for our benchmarks (X-axis). The execution time is broken down by the steps (shown as sub-bars) of the training algorithm. 
We observe that the speedups inversely correlate with the fraction of execution time of \step{2}. 
Note that \im's breakdown here is different from the {\em sequential} breakdown shown earlier in \figref{opportunity}. The 32-core baseline relatively increases \step{2}'s   fraction of time. 
The \ig configuration offers a typical and modest reduction in the three accelerated steps. Even on
the GPU, \step{2} does not see much improvement.
In contrast, \name  makes all the accelerated steps vanishingly small. \name's residual execution time is dominated by the unaccelerated \step{2}.

\putsubsec{isolation}{Isolating \name's optimizations}
\figref{isolation} isolates the speedup  contribution of \name's components (Y-axis, \im = 1) for our benchmarks (X-axis). 
The speedups are over \im (first bar).
The second bar {\em \name-no-opts} uses naive packing  to map bins to SRAMs (\secref{mapping}) and  no optimizations other than exploiting BU parallelism.
% It uses naive packing to map bins to SRAMs. 
The next bar shows the speedup due to group-by-field mapping (\secref{mapping}) which  shows improvements for the two benchmarks with categorical fields (\allstate and \flight). 
Recall that  the mapping is designed to improve datasets with categorical fields. For benchmarks without a single categorical field, naive packing achieves 
the same effect as the mapping because our SRAMs are sized to accommodate numerical fields.
Finally, the third bar shows the contribution of 
our redundant per-field, column-major representation which helps accelerate steps \rawstep{3} and \rawstep{5}. This optimization works best for benchmarks that already have high speedups. Effectively, speeding up \step{1} makes steps \rawstep{3} and \rawstep{5} take larger fractions of the residual execution time. 
As such, the impact of any optimization that targets steps \rawstep{3} and \rawstep{5} is magnified. 

The redundant representation, a software-only optimization, may be applied to steps \rawstep{3} and \rawstep{5} in the \ig and \im configurations. However,  \im and \ig see little benefits (under 4\%) because 
(1) \step{1} dominates the execution time for \im and \ig (see \figref{breakdown}), (2) \step{5} is compute-bound on \im and \ig, and (3) \step{3} does
 improve, but with minimal overall impact (Amdahl's Law effect).

\figput{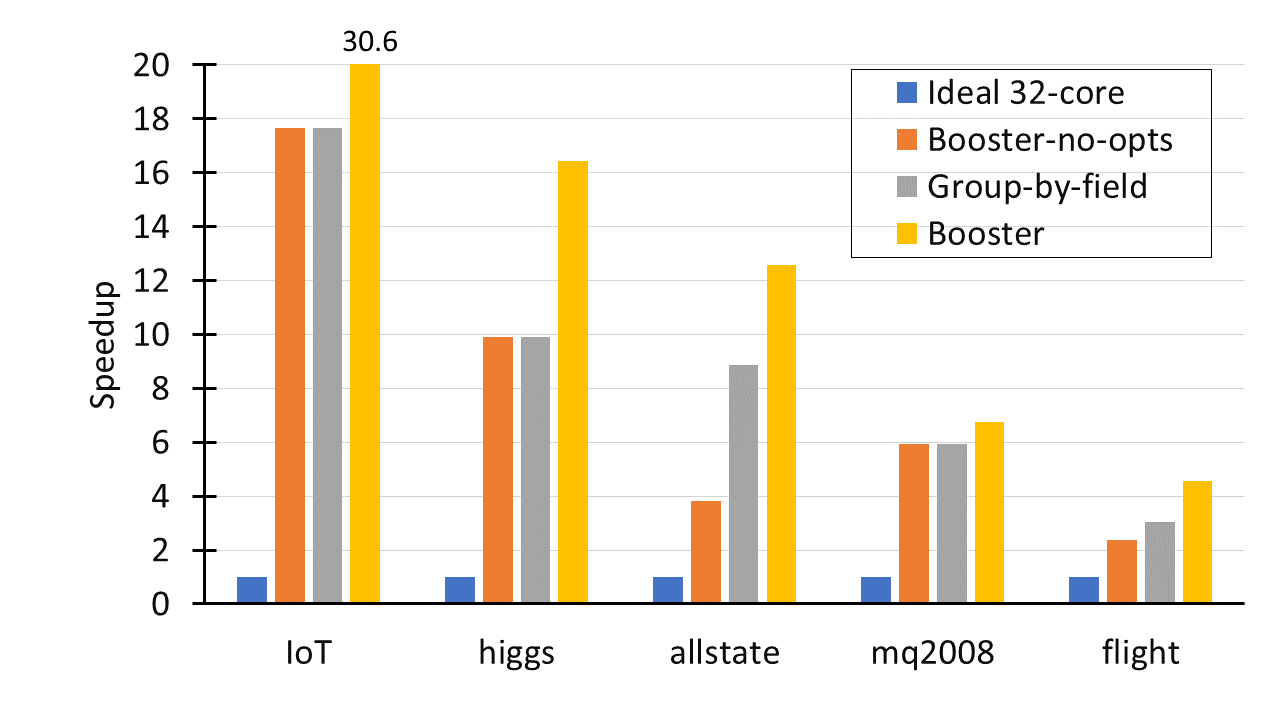}{}{Isolating the Impact of \name's  optimizations}

\putsubsec{energy}{Energy}

We show access energy comparisons for SRAM and DRAM separately as we cannot measure the relative proportions of the two components. However, because \name is strictly better in both SRAM energy {\em and} DRAM energy, \name is guaranteed to achieve lower total energy  regardless of the specific ratio of SRAM energy to DRAM energy. 

\figref{DRAMEnergy} compares  the  SRAM (\figref{DRAMEnergy}(a)) and  DRAM  
(\figref{DRAMEnergy}(b)) energy across the three configurations (\im, \ig, and \name). The results are averaged across all the benchmarks and normalized to that of \im. \ig's SRAM energy is higher because of the 32-way-banked, 96-KB Shared Memory  which incurs more energy than \im's 32-KB L1 D-cache. In contrast, \name accesses a smaller 2-KB SRAM,  resulting in lower SRAM energy.  For DRAM energy, \im and \ig are identical as they access the same set of blocks. In contrast, \name
has fewer  DRAM transfers via the per-field column-major format
%VJ XXX as well as  the cache-alignment tradeoff 
(\secref{microarch}). 

\figput[0.8\linewidth]{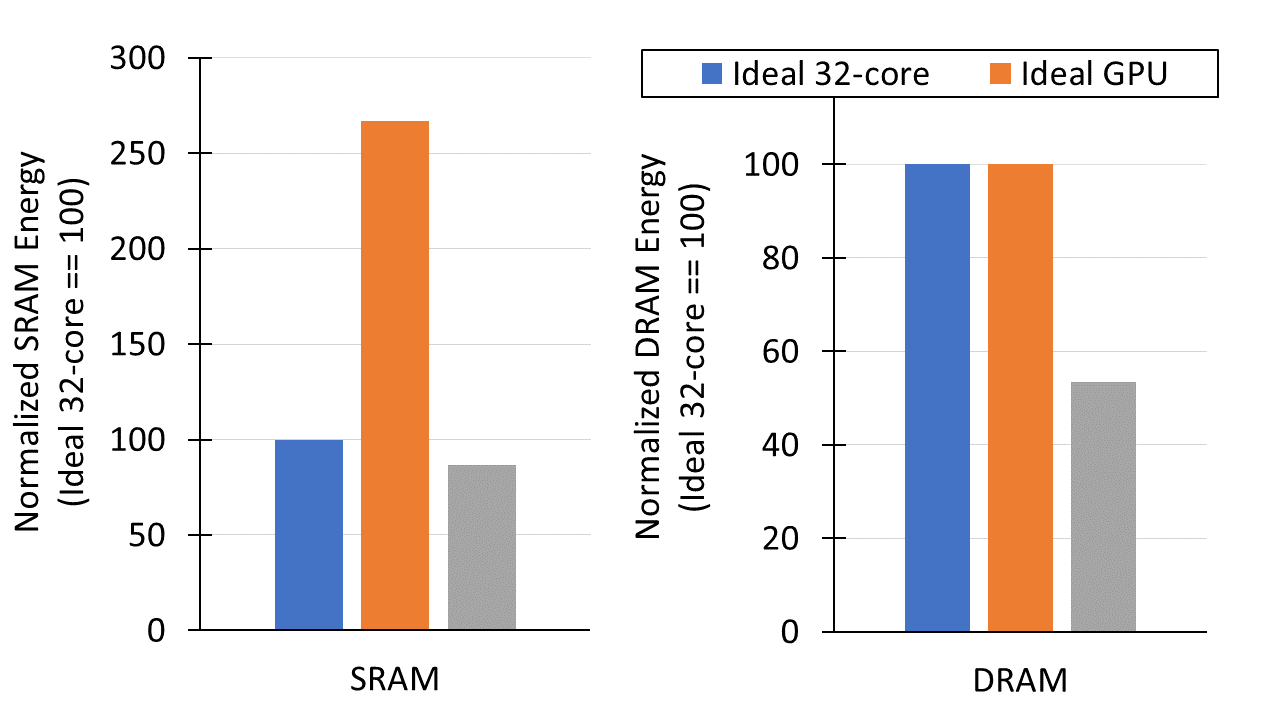}{}{\name's energy savings (a) SRAM and (b) DRAM}

 \putsubsec{sanity}{Validating the {\em Ideal} models}
 
\figput{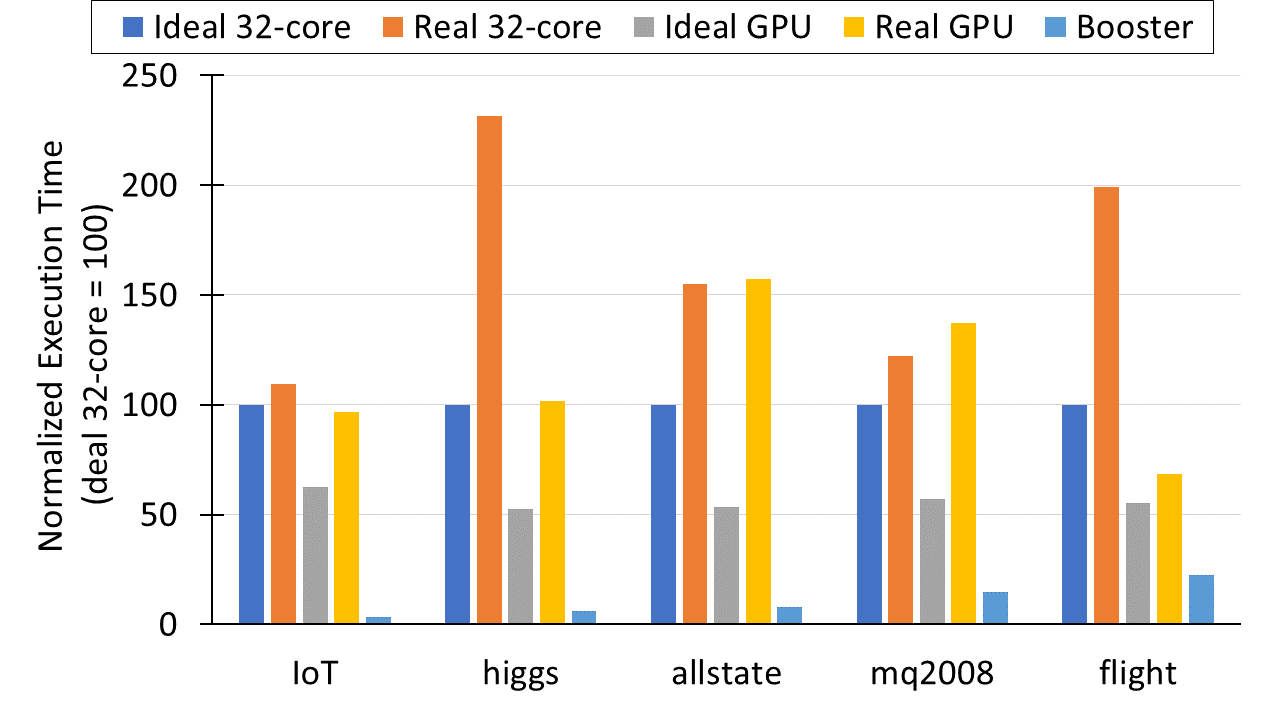}{}{Comparing the {\em Ideal} and {\em Real} configurations}
  
\figref{realcomp} compares the execution times on real hardware (32-core multicore and GPU) to those of \im and \ig for all our benchmarks (X-axis).  The last bar shows the execution time of \name. We use an Intel $5^{th}$ generation processor for the multicore and an Nvidia V100 for the GPU; both on Google Cloud. The execution times are normalized (Y-axis) to that of \im  (first bar), the same baseline used in all our speedup measurements. \figref{realcomp} confirms 
two important properties. 
First, the \im  execution time is always lower than that of the real 32-core  system.  
Similarly, the \ig configuration is always better-performing than the real GPU.

These results confirm the conservative nature of our comparisons;
\name's speedups are in comparison to aggressive, ideal 
baselines. 
Second, in comparing the GPU to the multicore configurations, our ideal model shows that \ig is always better-performing 
than \im. However, on real hardware, the results are mixed; 
GPU performance is worse than that of the multicore for two of the five benchmarks (\allstate and \mq). This result  confirms that the irregularity of the workload does indeed  limit the
speedups achievable on a GPU; thus strengthening the case 
for an accelerator. Finally, a direct comparison between \name and the real 32-core and GPU configurations, included for viewing convenience, confirms \name's speedups.

\putsubsec{scaleup}{Sensitivity to dataset size}
Given that commercial datasets are larger (\secref{method}) and that data is growing faster than Moore's law~\cite{datamoore},
it is important to understand how \name's performance is affected by growing dataset sizes.
To that end, we performed experiments by  scaling up (replicating) data sizes by a factor of 10. 
\figref{sensitivity} shows the performance comparison of the three hardware configurations (\im, \ig, and \name) for the scaled-up benchmarks (X-axis). 
The speedup of \ig over \im (Y-axis) remains modest ($<$ 2x) and similar to the speedups with the unscaled datasets. In contrast, \name's speedups are dramatically higher for the larger datasets. Compared with the unscaled datasets (\figref{main}), the speedup range improves from 4.6-30.6 to 9.8-61.5 with the scaled datasets; the geometric mean speedup increases from 11.4 to 27.9. 
{\em Every} benchmark achieves higher speedup with the scaled datasets than with the unscaled ones 
(\figref{main}); even \flight, the benchmark with the lowest speedup improves from
4.6 to 9.8 speedup.
This result highlights \name's strength that its performance will improve as data volume grows.

\figput{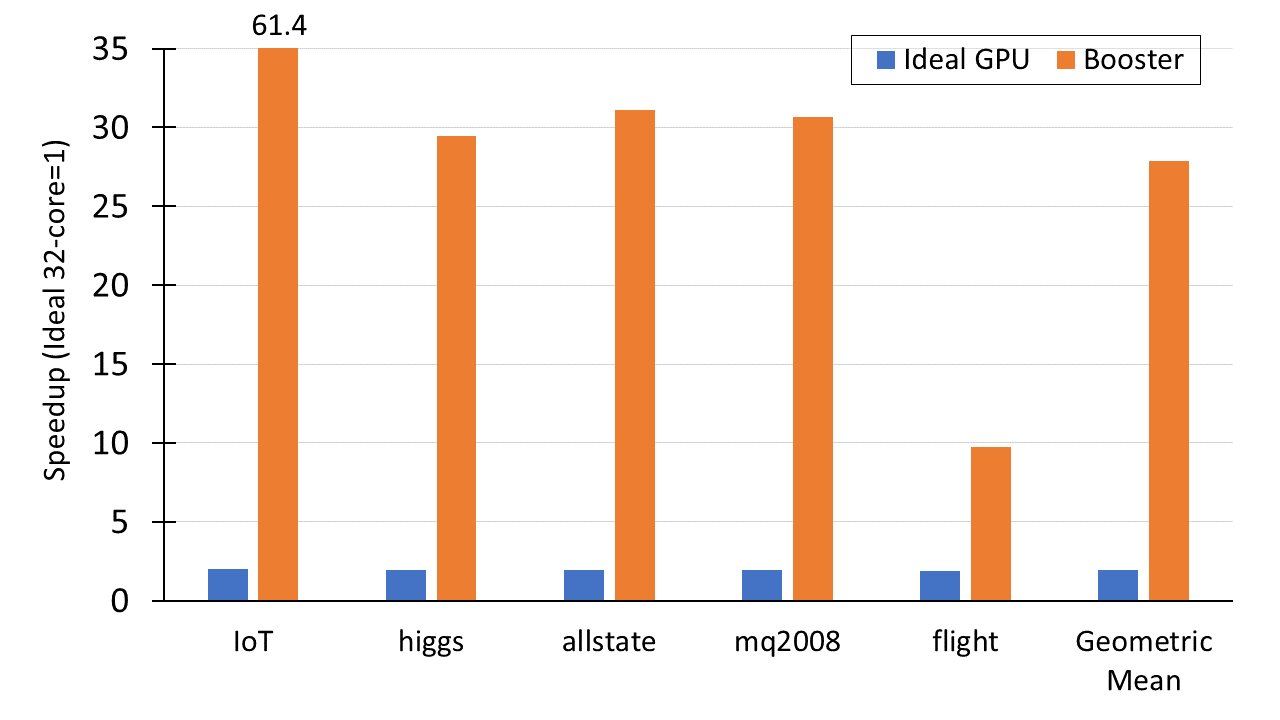}{}{Sensitivity to dataset size}

\putsubsec{asic-results}{RTL Validation and ASIC synthesis results}

\tabref{areapower} shows the area and power estimates for a 50-cluster \name chip with 64 BUs per cluster (using 40 nm technology). Our synthesis achieves 1-GHz clock speed. 
Almost half (55\%)  of \name's area goes to the SRAMs.
Although the aggregate SRAM capacity adds up to only 6.4 MB ($=3200*2$KB), the area is around 70\% larger than that of a 1-bank 6.4-MB SRAM array as \name uses the equivalent of 3200 banks for parallel access (the area would be much smaller for a recent technology node which is not available in FreePDK).
Because the static power dominates the dynamic power, \name's SRAM power is only around 59\% higher than that of the 1-bank case even though the latter makes only one access at a time while the former makes 3200 parallel accesses.
% as \name's small-array  accesses are 3200-way parallel.
While the  SRAM area is larger than the FPUs', the SRAM and FPUs dissipate nearly equal power, each accounting for around 41\% of the total power.

Because the  memory bandwidth on our FPGA platform was too
low to run meaningful benchmarks, we limited our FPGA runs to  verifying correctness.

\tabput{areapower}{
\begin{stripetabular}{ccc}
Component & Area (mm$^2$) & Power (W) \\
\hline
Control Logic & 8.4 & 4.3 \\ 
FPU & 18.4 & 9.5\\
SRAM & 33.1  & 9.4\\
\hline
Total & 60.0  & 23.2 \\
\end{stripetabular}
}
 {Area and power estimates for \name}

\putsubsec{Batch Inference}{Batch Inference}

We augmented our simulator for batch inference where each record traverses 500 trees. Although we did not implement inference in RTL, we did implement the key building block -- 
one-tree traversal -- in our design. We use the latency numbers from the one-tree traversal
model to simulate batch inference. Our simulation uses 3000 of the 3200 BUs to create 6 replicas of the 500 trees to improve inference throughput. 
\figref{batchinference} shows \name's speedup over \im (Y-axis) for batch inference over all the records in each benchmark (X-axis). Though our measurements show that \name achieves 45x mean speedup, the results are better understood as two distinct clustered behaviors.
Four of the five benchmarks with deep trees (as described in \secref{method}) behave similarly with 55.5x speedup over
over \im. The  outlier (\iot)  achieves 21.1x speedup because of its shallow trees. \name's absolute performance is unaffected as its performance depends
on the {\em maximum} depth across all trees, which is usually 6. In contrast,  \im's absolute performance improves for shallow trees due to reduced work (\secref{method}) which 
leads to \name's speedup being lower for \iot. 
Interestingly, \iot's shallow trees cause higher speedups for \name in training (\figref{main}), as opposed to inference. Shallow trees have less left-right sub-tree work-pruning in the histogram-binning \step{1} (\secref{xgboost-training}), so that \step{1} dominates the other steps, as explained in \secref{method}. By parallelizing this dominant step, \name  achieves high speedups in training.

\figput{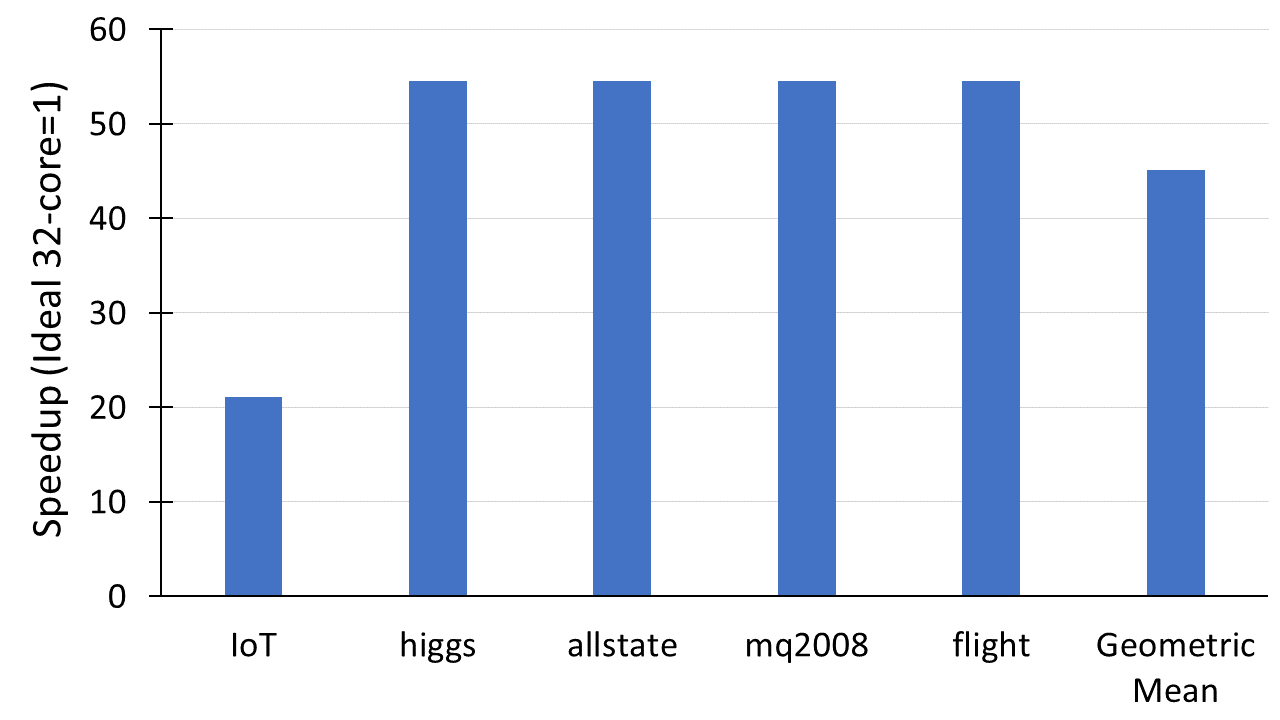}{}{\name's performance for Batch Inference}

\putsec{related}{Related work}

\paragraph*{Online Decision Trees}
Very Fast Decision Trees (VFDT)~\cite{vfdt} use a Hoeffding bound to guarantee that the decision points are asymptotically correct with respect to the offline variant.
Further, adaptive online decision trees  can handle concept drift \cite{Hulten2001, Bifet2009}. Others give theoretical guarantees for Adaptive Hoeffding Trees~\cite{Bifet2009}.

\paragraph*{Online Gradient Boosting}
The original online boosting algorithms maintain a set of weak online learners and updates all the weak online learners with every record using various weight updates~\cite{Oza2001, Oza2001a, Grabner2006, Leistner2009}.
Stochastic GB~\cite{Friedman2002}, where random subsets of samples are used for each tree rather than all samples, is shown to be more robust in certain cases. 
This approach can be seen as taking stochastic gradient descent steps in function space.

\name targets offline GB, the most widely-used variant. \name succeeds in accelerating the 
heavy work of training; processing millions of records over hours. 
If online training or inference processes one record or a small batch at a time, 
then {\em no} accelerator  (including \name) would  be justified. 

\paragraph*{ML Accelerators}
There are many DNN architecture proposals, including optimizations for compute~\cite{TPU, snowflake, ferdman-fpga2, alexnet, quantization, cnvlutin, bit-pragmatic, bit-tactical, scnn, sparten}, 
memory~\cite{dadiannao, pudiannao, shidiannao, EIE}, and 
reuse~\cite{eyeriss, fusion}. However, GB has not received much attention from architects.  

\putsec{concl}{Conclusion}

We proposed an accelerator --\name -- for training gradient boosting (GB) models.
\name's design is driven by the observation that the training for GB models 
is dominated by three key computational steps -- histogram binning, single-predicate evaluation, 
and one tree traversal -- that are performed on a large number of records. Together, 
these three steps account for 90\%-98\% of sequential training time. 
While there is abundant inter- and intra-record parallelism, traditional
multicores and GPUs are unable to parallelize the memory accesses which are irregular,  data-dependent, and to small data-structures. \name employs (1) a scalable sea-of-small-SRAMs approach to harness 
large-scale parallelism (e.g. 3200-way), and (2) a bandwidth-preserving mapping of data fields 
to SRAMs resulting in significantly high parallelism. Our simulations show that 
\name achieves 11.4x speedup and 6.4x speedup over an ideal 32-core multicore and an ideal GPU, respectively. Based on ASIC synthesis of FPGA-validated RTL using 45 nm technology, we estimate that a \name chip would occupy 60 mm$^2$ of area and  dissipate 23 W when operating at a 1-GHz clock speed.
Our results show that \name's speedups are higher for larger datasets; which is important given that 
data growth is surpassing Moore's law in recent years~\cite{datamoore}. Given these results, \name would be an attractive design for accelerating the highly-successful GB models.

\bibliographystyle{IEEEtranS}
\bibliography{local,cnn,gpu,david-other-bib,Mendeley-fixed,pnm}

% Generated by IEEEtranS.bst, version: 1.13 (2008/09/30)
\begin{thebibliography}{10}
\providecommand{\url}[1]{#1}
\csname url@samestyle\endcsname
\providecommand{\newblock}{\relax}
\providecommand{\bibinfo}[2]{#2}
\providecommand{\BIBentrySTDinterwordspacing}{\spaceskip=0pt\relax}
\providecommand{\BIBentryALTinterwordstretchfactor}{4}
\providecommand{\BIBentryALTinterwordspacing}{\spaceskip=\fontdimen2\font plus
\BIBentryALTinterwordstretchfactor\fontdimen3\font minus
  \fontdimen4\font\relax}
\providecommand{\BIBforeignlanguage}[2]{{%
\expandafter\ifx\csname l@#1\endcsname\relax
\typeout{** WARNING: IEEEtranS.bst: No hyphenation pattern has been}%
\typeout{** loaded for the language `#1'. Using the pattern for}%
\typeout{** the default language instead.}%
\else
\language=\csname l@#1\endcsname
\fi
#2}}
\providecommand{\BIBdecl}{\relax}
\BIBdecl

\bibitem{airline-data}
\BIBentryALTinterwordspacing
``{ Airline on-time performance}:,''
  \url{http://stat-computing.org/dataexpo/2009/}, 2009. [Online]. Available:
  \url{http://stat-computing.org/dataexpo/2009/}
\BIBentrySTDinterwordspacing

\bibitem{Allstate-data}
\BIBentryALTinterwordspacing
``{ Allstate claim data}:,''
  \url{https://www.kaggle.com/c/ClaimPredictionChallenge/}, 2013. [Online].
  Available: \url{https://www.kaggle.com/c/ClaimPredictionChallenge/}
\BIBentrySTDinterwordspacing

\bibitem{bit-pragmatic}
\BIBentryALTinterwordspacing
J.~Albericio, A.~Delmas, P.~Judd, S.~Sharify, G.~O'Leary, R.~Genov, and
  A.~Moshovos, ``Bit-pragmatic deep neural network computing,'' in
  \emph{Proceedings of the 50th Annual {IEEE/ACM} International Symposium on
  Microarchitecture, {MICRO} 2017, Cambridge, MA, USA, October 14-18, 2017},
  2017, pp. 382--394. [Online]. Available:
  \url{http://doi.acm.org/10.1145/3123939.3123982}
\BIBentrySTDinterwordspacing

\bibitem{cnvlutin}
\BIBentryALTinterwordspacing
J.~Albericio, P.~Judd, T.~H. Hetherington, T.~M. Aamodt, N.~D.~E. Jerger, and
  A.~Moshovos, ``Cnvlutin: Ineffectual-neuron-free deep neural network
  computing,'' in \emph{43rd {ACM/IEEE} Annual International Symposium on
  Computer Architecture, {ISCA} 2016, Seoul, South Korea, June 18-22, 2016},
  2016, pp. 1--13. [Online]. Available:
  \url{http://dx.doi.org/10.1109/ISCA.2016.11}
\BIBentrySTDinterwordspacing

\bibitem{fusion}
M.~Alwani, H.~Chen, M.~Ferdman, and P.~Milder, ``Fused-layer cnn
  accelerators,'' in \emph{49th Annual IEEE/ACM International Symposium on
  Microarchitecture (MICRO)}, 2016.

\bibitem{wassersteingan2017}
M.~Arjovsky, S.~Chintala, and L.~Bottou, ``{W}asserstein generative adversarial
  networks,'' in \emph{Proceedings of the 34th International Conference on
  Machine Learning}, ser. Proceedings of Machine Learning Research, D.~Precup
  and Y.~W. Teh, Eds., vol.~70.\hskip 1em plus 0.5em minus 0.4em\relax
  International Convention Centre, Sydney, Australia: PMLR, 06--11 Aug 2017,
  pp. 214--223.

\bibitem{awesome}
``{AWesome XGBoost},'' \url{https://github.com/dmlc/xgboost/tree/master/demo}.

\bibitem{cacti7.0}
\BIBentryALTinterwordspacing
R.~Balasubramonian, A.~B. Kahng, N.~Muralimanohar, A.~Shafiee, and V.~Srinivas,
  ``Cacti 7: New tools for interconnect exploration in innovative off-chip
  memories,'' \emph{ACM Trans. Archit. Code Optim.}, vol.~14, no.~2, Jun. 2017.
  [Online]. Available: \url{https://doi.org/10.1145/3085572}
\BIBentrySTDinterwordspacing

\bibitem{Higgs}
P.~Baldi, P.~Sadowski, and D.~Whiteson, ``Searching for exotic particles in
  high-energy physics with deep learning,'' \emph{Nature communications},
  vol.~5, p. 4308, 2014.

\bibitem{Bifet2009}
A.~Bifet, G.~Holmes, B.~Pfahringer, R.~Kirkby, and R.~Gavald{\`{a}}, ``New
  ensemble methods for evolving data streams,'' in \emph{Proceedings of the
  15th ACM SIGKDD international conference on Knowledge discovery and data
  mining - KDD '09}.\hskip 1em plus 0.5em minus 0.4em\relax New York, New York,
  USA: ACM Press, 2009, p. 139.

\bibitem{gpumem}
N.~{Chatterjee}, M.~{O’Connor}, D.~{Lee}, D.~R. {Johnson}, S.~W. {Keckler},
  M.~{Rhu}, and W.~J. {Dally}, ``Architecting an energy-efficient dram system
  for gpus,'' in \emph{2017 IEEE International Symposium on High Performance
  Computer Architecture (HPCA)}, 2017, pp. 73--84.

\bibitem{xgboost}
T.~Chen and C.~Guestrin, ``{XGBoost}: A scalable tree boosting system,'' in
  \emph{Proceedings of the 22nd ACM SIGKDD International Conference on
  Knowledge Discovery and Data Mining}.\hskip 1em plus 0.5em minus 0.4em\relax
  ACM, 2016, pp. 785--794.

\bibitem{diannao}
\BIBentryALTinterwordspacing
T.~Chen, Z.~Du, N.~Sun, J.~Wang, C.~Wu, Y.~Chen, and O.~Temam, ``Diannao: A
  small-footprint high-throughput accelerator for ubiquitous
  machine-learning,'' in \emph{Proceedings of the 19th International Conference
  on Architectural Support for Programming Languages and Operating Systems},
  ser. ASPLOS '14.\hskip 1em plus 0.5em minus 0.4em\relax New York, NY, USA:
  ACM, 2014, pp. 269--284. [Online]. Available:
  \url{http://doi.acm.org/10.1145/2541940.2541967}
\BIBentrySTDinterwordspacing

\bibitem{eyeriss}
Y.-H. Chen, T.~Krishna, J.~Emer, and V.~Sze, ``14.5 eyeriss: An
  energy-efficient reconfigurable accelerator for deep convolutional neural
  networks,'' in \emph{2016 IEEE International Solid-State Circuits Conference
  (ISSCC)}, Jan 2016, pp. 262--263.

\bibitem{dadiannao}
\BIBentryALTinterwordspacing
Y.~Chen, T.~Luo, S.~Liu, S.~Zhang, L.~He, J.~Wang, L.~Li, T.~Chen, Z.~Xu,
  N.~Sun, and O.~Temam, ``Dadiannao: A machine-learning supercomputer,'' in
  \emph{Proceedings of the 47th Annual IEEE/ACM International Symposium on
  Microarchitecture}, ser. MICRO-47.\hskip 1em plus 0.5em minus 0.4em\relax
  Washington, DC, USA: IEEE Computer Society, 2014, pp. 609--622. [Online].
  Available: \url{http://dx.doi.org/10.1109/MICRO.2014.58}
\BIBentrySTDinterwordspacing

\bibitem{seg1}
\BIBentryALTinterwordspacing
D.~C. Cire\c{s}an, A.~Giusti, L.~M. Gambardella, and J.~Schmidhuber, ``Deep
  neural networks segment neuronal membranes in electron microscopy images,''
  in \emph{Proceedings of the 25th International Conference on Neural
  Information Processing Systems - Volume 2}, ser. NIPS'12.\hskip 1em plus
  0.5em minus 0.4em\relax USA: Curran Associates Inc., 2012, pp. 2843--2851.
  [Online]. Available: \url{http://dl.acm.org/citation.cfm?id=2999325.2999452}
\BIBentrySTDinterwordspacing

\bibitem{bit-tactical}
\BIBentryALTinterwordspacing
A.~Delmas, P.~Judd, D.~M. Stuart, Z.~Poulos, M.~Mahmoud, S.~Sharify,
  M.~Nikolic, and A.~Moshovos, ``Bit-tactical: Exploiting ineffectual
  computations in convolutional neural networks: Which, why, and how,''
  \emph{CoRR}, vol. abs/1803.03688, 2018. [Online]. Available:
  \url{http://arxiv.org/abs/1803.03688}
\BIBentrySTDinterwordspacing

\bibitem{vfdt}
P.~Domingos and G.~Hulten, ``Mining high-speed data streams,'' in \emph{KDD},
  vol.~2, 2000, p.~4.

\bibitem{catboost}
A.~V. Dorogush, V.~Ershov, and A.~Gulin, ``{CatBoost}: gradient boosting with
  categorical features support,'' in \emph{Workshop on ML Systems at NIPS},
  2017.

\bibitem{shidiannao}
\BIBentryALTinterwordspacing
Z.~Du, R.~Fasthuber, T.~Chen, P.~Ienne, L.~Li, T.~Luo, X.~Feng, Y.~Chen, and
  O.~Temam, ``Shidiannao: Shifting vision processing closer to the sensor,'' in
  \emph{Proceedings of the 42Nd Annual International Symposium on Computer
  Architecture}, ser. ISCA '15.\hskip 1em plus 0.5em minus 0.4em\relax New
  York, NY, USA: ACM, 2015, pp. 92--104. [Online]. Available:
  \url{http://doi.acm.org/10.1145/2749469.2750389}
\BIBentrySTDinterwordspacing

\bibitem{seg2}
C.~{Farabet}, C.~{Couprie}, L.~{Najman}, and Y.~{LeCun}, ``Learning
  hierarchical features for scene labeling,'' \emph{IEEE Transactions on
  Pattern Analysis and Machine Intelligence}, vol.~35, no.~8, pp. 1915--1929,
  Aug 2013.

\bibitem{Friedman2001}
J.~H. Friedman, ``Greedy function approximation: A gradient boosting machine,''
  \emph{The Annals of Statistics}, vol.~29, no.~5, pp. 1189--1232, 2001.

\bibitem{Friedman2002}
J.~H. Friedman, ``Stochastic gradient boosting,'' \emph{Computational
  Statistics {\&} Data Analysis}, vol.~38, no.~4, pp. 367--378, feb 2002.

\bibitem{datamoore}
J.~Gantz and D.~Reinsel, ``Extracting value from chaos,'' \emph{IDC IView}, pp.
  1--12, 01 2011.

\bibitem{snowflake}
V.~Gokhale, A.~Zaidy, A.~X.~M. Chang, and E.~Culurciello, ``Snowflake: An
  efficient hardware accelerator for convolutional neural networks,'' in
  \emph{2017 IEEE International Symposium on Circuits and Systems (ISCAS)}, May
  2017, pp. 1--4.

\bibitem{sparten}
\BIBentryALTinterwordspacing
A.~Gondimalla, N.~Chesnut, M.~Thottethodi, and T.~N. Vijaykumar, ``Sparten: A
  sparse tensor accelerator for convolutional neural networks,'' in
  \emph{Proceedings of the 52Nd Annual IEEE/ACM International Symposium on
  Microarchitecture}, ser. MICRO '52.\hskip 1em plus 0.5em minus 0.4em\relax
  New York, NY, USA: ACM, 2019, pp. 151--165. [Online]. Available:
  \url{http://doi.acm.org/10.1145/3352460.3358291}
\BIBentrySTDinterwordspacing

\bibitem{Goodfellow2014}
I.~Goodfellow, J.~Pouget-Abadie, M.~Mirza, B.~Xu, D.~Warde-Farley, S.~Ozair,
  A.~Courville, and Y.~Bengio, ``Generative adversarial nets,'' in \emph{Neural
  Information Processing Systems (NIPS)}, 2014.

\bibitem{Grabner2006}
H.~Grabner and H.~Bischof, ``On-line boosting and vision,'' in \emph{IEEE
  Computer Society Conference on Computer Vision and Pattern Recognition
  (CVPR'06)}, vol.~1.\hskip 1em plus 0.5em minus 0.4em\relax IEEE, 2006, pp.
  260--267.

\bibitem{EIE}
S.~Han, X.~Liu, H.~Mao, J.~Pu, A.~Pedram, M.~A. Horowitz, and W.~J. Dally,
  ``Eie: Efficient inference engine on compressed deep neural network,'' in
  \emph{2016 ACM/IEEE 43rd Annual International Symposium on Computer
  Architecture (ISCA)}, June 2016, pp. 243--254.

\bibitem{resnet}
\BIBentryALTinterwordspacing
K.~He, X.~Zhang, S.~Ren, and J.~Sun, ``Deep residual learning for image
  recognition,'' \emph{CoRR}, vol. abs/1512.03385, 2015. [Online]. Available:
  \url{http://arxiv.org/abs/1512.03385}
\BIBentrySTDinterwordspacing

\bibitem{facebook-adclick}
\BIBentryALTinterwordspacing
X.~He, J.~Pan, O.~Jin, T.~Xu, B.~Liu, T.~Xu, Y.~Shi, A.~Atallah, R.~Herbrich,
  S.~Bowers, and J.~Q.~n. Candela, ``Practical lessons from predicting clicks
  on ads at facebook,'' in \emph{Proceedings of the Eighth International
  Workshop on Data Mining for Online Advertising}, ser. ADKDD’14.\hskip 1em
  plus 0.5em minus 0.4em\relax New York, NY, USA: Association for Computing
  Machinery, 2014, p. 1–9. [Online]. Available:
  \url{https://doi.org/10.1145/2648584.2648589}
\BIBentrySTDinterwordspacing

\bibitem{hgdata}
``{HG Insights},'' \url{(https://discovery.hgdata.com/product/xgboost)}.

\bibitem{Hulten2001}
G.~Hulten, L.~Spencer, and P.~Domingos, ``Mining time-changing data streams,''
  in \emph{Proceedings of the seventh ACM SIGKDD international conference on
  Knowledge discovery and data mining - KDD '01}.\hskip 1em plus 0.5em minus
  0.4em\relax New York, New York, USA: ACM Press, 2001, pp. 97--106.

\bibitem{isola2017image}
P.~Isola, J.-Y. Zhu, T.~Zhou, and A.~A. Efros, ``Image-to-image translation
  with conditional adversarial networks,'' in \emph{Proceedings of the IEEE
  conference on computer vision and pattern recognition}, 2017, pp. 1125--1134.

\bibitem{TPU}
\BIBentryALTinterwordspacing
N.~P. Jouppi, C.~Young, N.~Patil, D.~Patterson, G.~Agrawal, R.~Bajwa, S.~Bates,
  S.~Bhatia, N.~Boden, A.~Borchers, R.~Boyle, P.-l. Cantin, C.~Chao, C.~Clark,
  J.~Coriell, M.~Daley, M.~Dau, J.~Dean, B.~Gelb, T.~V. Ghaemmaghami,
  R.~Gottipati, W.~Gulland, R.~Hagmann, C.~R. Ho, D.~Hogberg, J.~Hu, R.~Hundt,
  D.~Hurt, J.~Ibarz, A.~Jaffey, A.~Jaworski, A.~Kaplan, H.~Khaitan,
  D.~Killebrew, A.~Koch, N.~Kumar, S.~Lacy, J.~Laudon, J.~Law, D.~Le, C.~Leary,
  Z.~Liu, K.~Lucke, A.~Lundin, G.~MacKean, A.~Maggiore, M.~Mahony, K.~Miller,
  R.~Nagarajan, R.~Narayanaswami, R.~Ni, K.~Nix, T.~Norrie, M.~Omernick,
  N.~Penukonda, A.~Phelps, J.~Ross, M.~Ross, A.~Salek, E.~Samadiani, C.~Severn,
  G.~Sizikov, M.~Snelham, J.~Souter, D.~Steinberg, A.~Swing, M.~Tan,
  G.~Thorson, B.~Tian, H.~Toma, E.~Tuttle, V.~Vasudevan, R.~Walter, W.~Wang,
  E.~Wilcox, and D.~H. Yoon, ``In-datacenter performance analysis of a tensor
  processing unit,'' in \emph{Proceedings of the 44th Annual International
  Symposium on Computer Architecture}, ser. ISCA '17.\hskip 1em plus 0.5em
  minus 0.4em\relax New York, NY, USA: ACM, 2017, pp. 1--12. [Online].
  Available: \url{http://doi.acm.org/10.1145/3079856.3080246}
\BIBentrySTDinterwordspacing

\bibitem{lightgbm}
G.~Ke, Q.~Meng, T.~Finley, T.~Wang, W.~Chen, W.~Ma, Q.~Ye, and T.-Y. Liu,
  ``{LightGBM}: A highly efficient gradient boosting decision tree,'' in
  \emph{Neural Information Processing Systems (NeurIPS)}, 2017, pp. 3146--3154.

\bibitem{fastbdt}
T.~Keck, ``Fastbdt: A speed-optimized and cache-friendly implementation of
  stochastic gradient-boosted decision trees for multivariate classification,''
  \emph{arXiv preprint arXiv:1609.06119}, 2016.

\bibitem{alexnet}
\BIBentryALTinterwordspacing
A.~Krizhevsky, I.~Sutskever, and G.~E. Hinton, ``Imagenet classification with
  deep convolutional neural networks,'' in \emph{Advances in Neural Information
  Processing Systems 25}, F.~Pereira, C.~J.~C. Burges, L.~Bottou, and K.~Q.
  Weinberger, Eds.\hskip 1em plus 0.5em minus 0.4em\relax Curran Associates,
  Inc., 2012, pp. 1097--1105. [Online]. Available:
  \url{http://papers.nips.cc/paper/4824-imagenet-classification-with-deep-convolutional-neural-networks.pdf}
\BIBentrySTDinterwordspacing

\bibitem{Leistner2009}
C.~Leistner, A.~Saffari, P.~M. Roth, and H.~Bischof, ``On robustness of on-line
  boosting - a competitive study,'' in \emph{2009 IEEE 12th International
  Conference on Computer Vision Workshops, ICCV Workshops}.\hskip 1em plus
  0.5em minus 0.4em\relax IEEE, sep 2009, pp. 1362--1369.

\bibitem{quantization}
\BIBentryALTinterwordspacing
D.~D. Lin, S.~S. Talathi, and V.~S. Annapureddy, ``Fixed point quantization of
  deep convolutional networks,'' in \emph{Proceedings of the 33rd International
  Conference on International Conference on Machine Learning - Volume 48}, ser.
  ICML'16.\hskip 1em plus 0.5em minus 0.4em\relax JMLR.org, 2016, pp.
  2849--2858. [Online]. Available:
  \url{http://dl.acm.org/citation.cfm?id=3045390.3045690}
\BIBentrySTDinterwordspacing

\bibitem{bing-adclick}
\BIBentryALTinterwordspacing
X.~Ling, W.~Deng, C.~Gu, H.~Zhou, C.~Li, and F.~Sun, ``Model ensemble for click
  prediction in bing search ads,'' in \emph{Proceedings of the 26th
  International Conference on World Wide Web Companion}, ser. WWW ’17
  Companion.\hskip 1em plus 0.5em minus 0.4em\relax Republic and Canton of
  Geneva, CHE: International World Wide Web Conferences Steering Committee,
  2017, p. 689–698. [Online]. Available:
  \url{https://doi.org/10.1145/3041021.3054192}
\BIBentrySTDinterwordspacing

\bibitem{pudiannao}
\BIBentryALTinterwordspacing
D.~Liu, T.~Chen, S.~Liu, J.~Zhou, S.~Zhou, O.~Teman, X.~Feng, X.~Zhou, and
  Y.~Chen, ``Pudiannao: A polyvalent machine learning accelerator,'' in
  \emph{Proceedings of the Twentieth International Conference on Architectural
  Support for Programming Languages and Operating Systems}, ser. ASPLOS
  '15.\hskip 1em plus 0.5em minus 0.4em\relax New York, NY, USA: ACM, 2015, pp.
  369--381. [Online]. Available:
  \url{http://doi.acm.org/10.1145/2694344.2694358}
\BIBentrySTDinterwordspacing

\bibitem{IOT-data}
Y.~Meidan, M.~Bohadana, Y.~Mathov, Y.~Mirsky, A.~Shabtai, D.~Breitenbacher, and
  Y.~Elovici, ``N-baiot—network-based detection of iot botnet attacks using
  deep autoencoders,'' \emph{IEEE Pervasive Computing}, vol.~17, no.~3, pp.
  12--22, 2018.

\bibitem{freepdk45}
NCSU, ``Freepdk45,''
  \url{https://www.eda.ncsu.edu/wiki/FreePDK45:Contents#Current_Version }.

\bibitem{Oza2001a}
N.~C. Oza and S.~Russell, ``Experimental comparisons of online and batch
  versions of bagging and boosting,'' in \emph{Proceedings of the seventh ACM
  SIGKDD international conference on Knowledge discovery and data mining
  (KDD)}.\hskip 1em plus 0.5em minus 0.4em\relax New York, New York, USA: ACM
  Press, 2001, pp. 359--364.

\bibitem{Oza2001}
N.~C. Oza and S.~Russell, ``Online bagging and boosting,'' in \emph{Eighth
  International Workshop on Artificial Intelligence and Statistics}, 2001.

\bibitem{Flight-data}
S.~Pafka, ``{ Flight delay data }:,'' \url{
  https://github.com/szilard/benchm-ml#data}, 2016.

\bibitem{scnn}
\BIBentryALTinterwordspacing
A.~Parashar, M.~Rhu, A.~Mukkara, A.~Puglielli, R.~Venkatesan, B.~Khailany,
  J.~Emer, S.~W. Keckler, and W.~J. Dally, ``Scnn: An accelerator for
  compressed-sparse convolutional neural networks,'' in \emph{Proceedings of
  the 44th Annual International Symposium on Computer Architecture}, ser. ISCA
  '17.\hskip 1em plus 0.5em minus 0.4em\relax New York, NY, USA: ACM, 2017, pp.
  27--40. [Online]. Available: \url{http://doi.acm.org/10.1145/3079856.3080254}
\BIBentrySTDinterwordspacing

\bibitem{Mq2008}
T.~Qin and T.-Y. Liu, ``Introducing letor 4.0 datasets,'' \emph{arXiv preprint
  arXiv:1306.2597}, 2013.

\bibitem{dramsim}
P.~{Rosenfeld}, E.~{Cooper-Balis}, and B.~{Jacob}, ``Dramsim2: A cycle accurate
  memory system simulator,'' \emph{IEEE Computer Architecture Letters},
  vol.~10, no.~1, pp. 16--19, 2011.

\bibitem{vggnet}
\BIBentryALTinterwordspacing
O.~Russakovsky, J.~Deng, H.~Su, J.~Krause, S.~Satheesh, S.~Ma, Z.~Huang,
  A.~Karpathy, A.~Khosla, M.~S. Bernstein, A.~C. Berg, and F.~Li, ``Imagenet
  large scale visual recognition challenge,'' \emph{CoRR}, vol. abs/1409.0575,
  2014. [Online]. Available: \url{http://arxiv.org/abs/1409.0575}
\BIBentrySTDinterwordspacing

\bibitem{GB-Sadasue}
T.~{Sadasue} and T.~{Isshiki}, ``Scalable full hardware logic architecture for
  gradient boosted tree training,'' in \emph{2020 IEEE 28th Annual
  International Symposium on Field-Programmable Custom Computing Machines
  (FCCM)}, 2020, pp. 234--234.

\bibitem{Schapire2012}
R.~E. Schapire and Y.~Freund, \emph{Boosting: Foundations and
  Algorithms}.\hskip 1em plus 0.5em minus 0.4em\relax Cambridge, MA: The MIT
  Press, 2012.

\bibitem{WESAD}
\BIBentryALTinterwordspacing
P.~Schmidt, A.~Reiss, R.~Duerichen, C.~Marberger, and K.~Van~Laerhoven,
  ``Introducing wesad, a multimodal dataset for wearable stress and affect
  detection,'' in \emph{Proceedings of the 20th ACM International Conference on
  Multimodal Interaction}, ser. ICMI ’18.\hskip 1em plus 0.5em minus
  0.4em\relax New York, NY, USA: Association for Computing Machinery, 2018, p.
  400–408. [Online]. Available: \url{https://doi.org/10.1145/3242969.3242985}
\BIBentrySTDinterwordspacing

\bibitem{ferdman-fpga2}
Y.~Shen, M.~Ferdman, and P.~Milder, ``{Escher}: A {CNN} accelerator with
  flexible buffering to minimize off-chip transfer,'' in \emph{25th IEEE
  International Symposium on Field-Programmable Custom Computing Machines
  ({FCCM})}, 2017.

\bibitem{inception}
\BIBentryALTinterwordspacing
C.~Szegedy, W.~Liu, Y.~Jia, P.~Sermanet, S.~Reed, D.~Anguelov, D.~Erhan,
  V.~Vanhoucke, and A.~Rabinovich, ``Going deeper with convolutions,'' in
  \emph{Computer Vision and Pattern Recognition (CVPR)}, 2015. [Online].
  Available: \url{http://arxiv.org/abs/1409.4842}
\BIBentrySTDinterwordspacing

\bibitem{GB-Tanaka}
T.~Tanaka, R.~Kasahara, and D.~Kobayashi, ``Efficient logic architecture in
  training gradient boosting decision tree for high-performance and edge
  computing,'' 2018.

\bibitem{terabyteclicklog}
``{Terabyte Click Log},''
  \url{(https://labs.criteo.com/2013/12/download-terabyte-click-logs-2/)}.

\bibitem{terabyteclicklog-runtime}
``{LightGBM Experiments},''
  \url{(https://lightgbm.readthedocs.io/en/latest/Experiments.html)}.

\bibitem{ThunderGBM}
Z.~{Wen}, J.~{Shi}, B.~{He}, J.~{Chen}, K.~{Ramamohanarao}, and Q.~{Li},
  ``Exploiting gpus for efficient gradient boosting decision tree training,''
  \emph{IEEE Transactions on Parallel and Distributed Systems}, vol.~30,
  no.~12, pp. 2706--2717, Dec 2019.

\bibitem{xgboost-wikipedia}
``{Wikipedia},'' \url{(https://en.wikipedia.org/wiki/XGBoost)}.

\bibitem{zhu2017unpaired}
J.-Y. Zhu, T.~Park, P.~Isola, and A.~A. Efros, ``Unpaired image-to-image
  translation using cycle-consistent adversarial networks,'' in
  \emph{Proceedings of the IEEE international conference on computer vision},
  2017, pp. 2223--2232.

\end{thebibliography}

\end{document}